\documentclass[9pt, reprint,amsmath,amssymb,apsm,siunitx,twocolumn]{revtex4-2}
\usepackage{float}
\makeatletter
\let\newfloat\newfloat@ltx
\makeatother
\usepackage{algorithm}
\usepackage{algpseudocode}
\usepackage[utf8]{inputenc}
\usepackage{natbib}
\usepackage{color}
\usepackage{wasysym}
\usepackage{graphicx}% Include figure files
\usepackage{adjustbox} %Resizes tables 
\usepackage{dcolumn}% Align table columns on decimal point
\usepackage{breqn}
\usepackage{physics}
\usepackage[toc,page]{appendix}%Adds appendix 
\usepackage{tabularx}
\usepackage{steinmetz}
\usepackage{makecell, booktabs, multirow}

\usepackage{lipsum, fancyvrb}
\usepackage[caption=false]{subfig} 
\makeatletter
\renewcommand*{\fnum@figure}{{\normalfont\bfseries \figurename~\thefigure}}
\renewcommand*{\@caption@fignum@sep}{\textbf{ : }}
\usepackage{fancyhdr}

\makeatother

\begin{document}

\title{Fast laser field reconstruction method based on a Gerchberg-Saxton algorithm with mode decomposition }

\author{I. Moulanier$^{1,*}$$,$ L. T. Dickson$^1$$,$ F. Massimo$^1$$,$ G. Maynard$^1$$,$ B. Cros$^1$}

\affiliation{1 LPGP$,$ CNRS$,$ Universit$\text{é}$ Paris Saclay$,$ 91405 Orsay$,$ France\\
* Corresponding author: ioaquin.moulanier@universite-paris-saclay.fr}

\begin{abstract}
Knowledge of the electric field of femtosecond, high intensity laser pulses is of paramount importance to study the interaction of this class of lasers with matter.
A novel, hybrid method to reconstruct the laser field from fluence measurements in the transverse plane at multiple positions along the propagation axis is presented, combining a Hermite-Gauss modes decomposition and elements of the Gerchberg-Saxton algorithm. 
The proposed Gerchberg-Saxton algorithm with modes decomposition (GSA-MD) takes into account the pointing instabilities of high intensity laser systems by tuning the centers of the HG modes.
Furthermore, it quickly builds a field description by progressively increasing the number of modes and thus the accuracy of the field reconstruction.
The results of field reconstruction using the GSA-MD are shown to be in excellent agreement with experimental measurements from two different high-peak power laser facilities.
\end{abstract}

\maketitle

\section{Introduction}
High intensity femtosecond laser pulses generated through chirped pulse amplification \cite{Strickland1985} are frequently affected by intensity and wavefront aberrations and fluctuations originating from multiple causes, e.g. thermal effects or imperfections of optical systems, inhomogeneities in the amplifying crystals' doping \cite{Ranc2000}, or air turbulence \cite{Yoon2021}. In addition, phase instabilities may result in pointing fluctuations and lack of symmetry of energy distribution in the focal volume \cite{Dickson2022, POP_IM}.

\begin{figure}[ht!]
\centering
\includegraphics[scale=0.35]{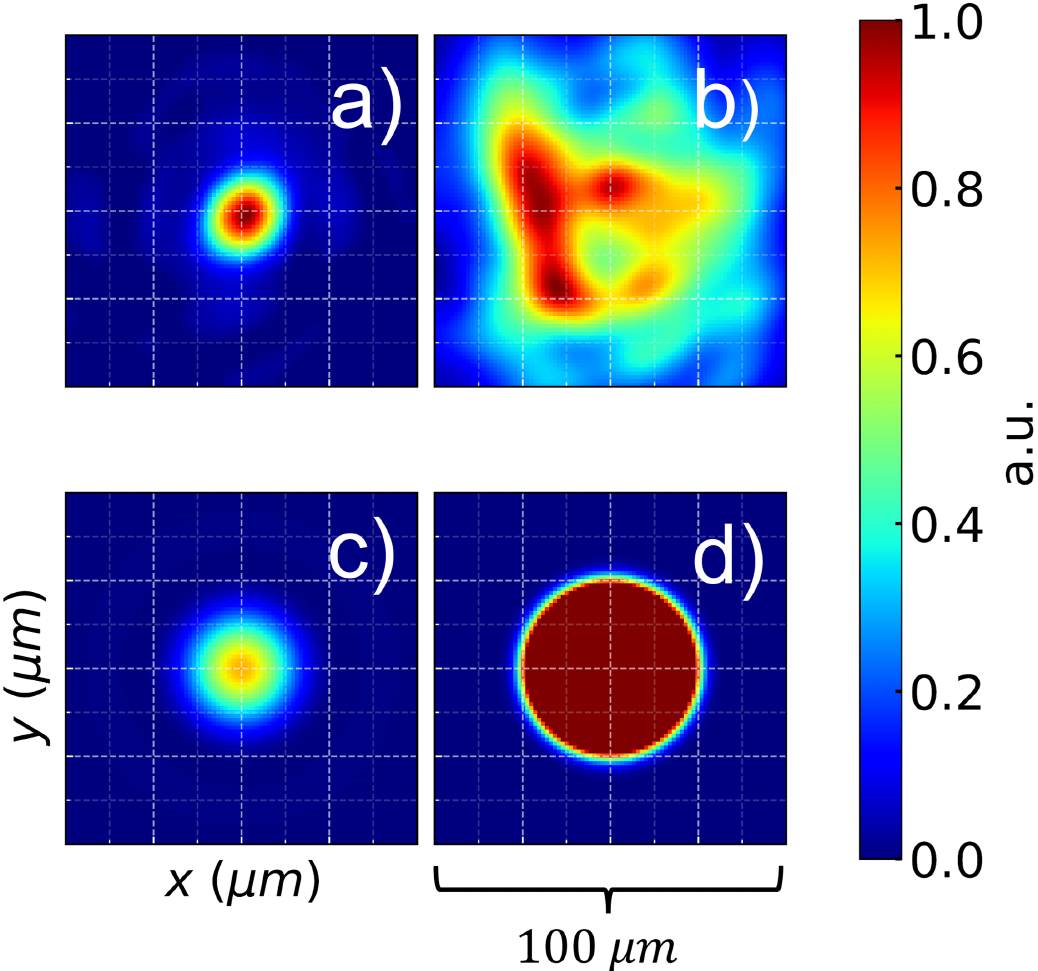}
\caption{Top row: an example of high intensity laser fluence map measured in an experiment : (a) at focus - (b) at $1500$ $\mu m$ from the focal plane. Bottom row: fluence corresponding to a 10th order Flattened Gaussian laser field distribution with the same energy : (c) at focus - (d) at $1500$ $\mu m$ from the focal plane. At each position, the maximum fluence  has been normalized to 1.}
\label{fig:asymmetry}
\end{figure}

An illustrative example of transverse asymmetry is shown in Fig. \ref{fig:asymmetry}, where the measured fluence of a 23 TW, 38 fs laser pulse on the top row is compared to the calculated fluence of a cylindrically symmetric flattened Gaussian transverse laser field distribution \cite{Santarsiero1997} in the bottom row.
Figure \ref{fig:asymmetry}a) shows that even in the focal plane, the transverse fluence distribution is asymmetric. At a larger distance from the focal plane (Fig. \ref{fig:asymmetry}b), the imperfections in the fluence distribution become even more pronounced.

In addition, spatio-temporal coupling (STC) of phase aberrations \cite{Akturk2010,Jeandet2022} reduce the quality of ultra-short high intensity laser pulses by increasing their duration and decreasing their peak intensity \cite{Bourassin-Bouchet2011,Li2017,Li2018,Jeandet2022}. Due to the nonlinear nature of the interaction of high intensity lasers with plasmas, these imperfections can decrease the laser peak intensity in the focal plane  \cite{Fourmaux2008} and degrade its symmetry \cite{Zemzemi_2020}, leading to lower performances e.g. for high harmonic generation \cite{Wodzinski2020} or laser wakefield acceleration (LWFA) \cite{Beaurepaire2015,ferri2016effect,Dickson2022}. 
These imperfections need to be mitigated in future applications of high intensity lasers like strong field quantum electrodynamics \cite{DiPiazza2012,Blackburn2020}, where reaching ultra high intensities and stable focusing is crucial. The study (and correction \cite{Fourmaux2008,Yoon2021}) of transverse aberrations requires intensity and wavefront measurements. However, measuring the wavefront of an intense, short laser pulse \cite{Wang2014} is more difficult than measuring the transverse laser fluence. For this reason, numerical methods to reconstruct the laser pulse wavefronts from fluence measurements are of paramount importance.
%Applications of phase reconstruction: Inertial Confinement Fusion \cite{Liu2002,Neauport2003,Li2017,Gao2020}\\

An important class of algorithms to retrieve the laser field from fluence measurements in two (or more) transverse planes along the propagation axis  originates from the Gerchberg-Saxton algorithm (GSA) \cite{gerchberg1972practical,Yang1994,Misell1973,Fienup1982,Zhou2019}.
In the basic formulation of the algorithm \cite{gerchberg1972practical}, the fluences measured at plane positions $z_0$ and $z_1$ (assuming a laser propagation along the $z$ direction) are used to build a progressively more accurate estimate of the field phase at $z_0$, starting from a random phase distribution at $z_0$. 
The algorithm, which performs an alternating field reconstruction at the two planes, is repeated until a stopping criterion is met, e.g. reaching a certain number of iterations, or reaching a certain value of a chosen reconstruction error metric. 
In the original article presenting the GSA it is shown that this error will decrease with the number of iterations \cite{gerchberg1972practical}, however the rate of convergence is undefined. Modifications of the original algorithm can yield a quicker convergence \cite{Fienup1982}.
Another important class of algorithms aims at reconstructing the field through an expansion with basis functions, e.g. the 
Nijboer-Zernike basis \cite{Antonello2015,Doelman2018,Miao22,Weise2023}. The algorithms in \cite{Santarsiero1999, Alieva2002} use an expansion in Hermite-Gauss (HG) modes to reconstruct the HG modal content of a signal, under some assumptions (e.g. finite modal content, knowledge of the HG modes spot sizes).
Since the analytical expression of the basis functions is known, these methods are often quicker than those derived from the GSA.

%To the best of the authors' knowledge, the algorithms belonging to this second class have not been used with TW-class or PW-class femtosecond laser pulses. This can be due to multiple reasons, for example the unknown modal content (as number of modes, and their spot sizes) of a realistic high-intensity laser field profile with this kind of laser pulses; besides, the fluence measurements at different planes come from different shots with shot-to-shot phase and pointing instabilities.
%These conditions would make the application of the cited algorithms using an expansion in function bases particularly challenging.

In this article, a hybrid field reconstruction method, called in the following Gerchberg-Saxton Algorithm with Modes Decomposition (GSA-MD), is presented. The GSA-MD combines field expansion in HG modes and some concepts of GSA algorithms, i.e. an iterative procedure, the phase extraction of the propagated field and the combination of this phase with the field amplitude measured at different planes.
%Whereas the original GSA and the algorithm in \cite{Miao22} are limited to fluence measurements in only two planes, the GSA-MD calculates the field without this restriction, as in the 3D GSA version presented in \cite{Zhou2019}.
\textcolor{black}{Whereas the original GSA \cite{gerchberg1972practical} and e.g. the algorithm in \cite{Miao22} are limited to fluence measurements in only two planes, 3D GSA variants in multi-plane propagation problems have been demonstrated \cite{ivanov1992phase, chessa1999phase, Zhou2019}. The GSA-MD can be used to reconstruct the electric field without any restriction on the number of planes.}
The GSA-MD addresses the uncertainty resulting from pointing instabilities affecting the fluence measurements by separating two problems: I) the field reconstruction, i.e. finding the coefficients in its HG modes decomposition, and II) the optimization of the choice of HG modes centers used in I) to reduce the reconstruction error.

Compared to previous versions of the GSA, the GSA-MD has several additional advantages.
As discussed in the following section, the conceptual separation of the two problems I) and II) avoids a direct, computationally prohibitive field reconstruction procedure. It will be shown that, in cases of interest, the number of unknowns in the proposed method is considerably lower than the number of unknowns with a classic GSA.
Other advantages of the GSA-MD are related to its flexibility. 
For example, depending on the type of  field distributions, different techniques can be independently used to solve the two mentioned problems, e.g other analytically known paraxial basis functions instead of the Hermite-Gauss modes can be used to address problem I), and various optimization algorithms can be used to address problem II).
Furthermore, using an expansion in HG modes in problem I) allows to choose the number of modes. It will be shown that this degree of freedom allows to perform a quick estimate of the HG modes coefficients with a low number of modes. This estimate can be subsequently refined using a higher number of modes, yielding an overall quicker field reconstruction. 
Finally, as it will be discussed in the following, the most computationally expensive steps of the GSA-MD can in principle be easily parallelized, since they act on independent HG modes. This is an advantage compared to the classic GSA, where the corresponding propagation steps are performed with Fourier transforms  \cite{gerchberg1972practical}, which are not easily parallelized.

%Furthermore, as will be shown in the Appendix, the proposed method can be generalized, increasing its convergence rate or improving the field reconstruction accuracy in a chosen plane.

An example application of the GSA is LWFA \cite{TajimaDawson1979,Esarey2009}, where it has been shown that including the GSA-reconstructed laser field in Particle in Cell simulations \cite{BirdsallLangdon2004} can greatly improve the agreement between simulations and measurements in the highly nonlinear regimes of laser-plasma interaction inherent to this field \cite{Beaurepaire2015,ferri2016effect}. 
The application of the proposed GSA-MD to LWFA modeling has been first presented in \cite{POP_IM}. In that reference it is shown that including a laser field reconstruction obtained with the GSA-MD in LWFA simulations considerably improves the agreement between simulated and measured energy-divergence electron spectra, compared to using simulations with ideal laser field distributions (as those in the bottom row in Fig. \ref{fig:asymmetry}). Here, a more detailed description of the field reconstruction method used  is reported. The GSA-MD in this article neglects the STC that may be present in the laser field. Future work may address the reconstruction of the laser field taking into account also these spatio-temporal imperfections.

The article is organised as follows. In the second section, an overview of the GSA-MD, including the description of the solutions to problems I) and II), is presented. In the third section, the results of the GSA-MD on two data-sets are shown. These two data-sets are made of fluence measurements at multiple planes performed at the Lund Laser Centre (LLC) and Apollon laser system in 2021. 

\begin{figure}[ht!]
\centering
\includegraphics[width=0.9\linewidth]{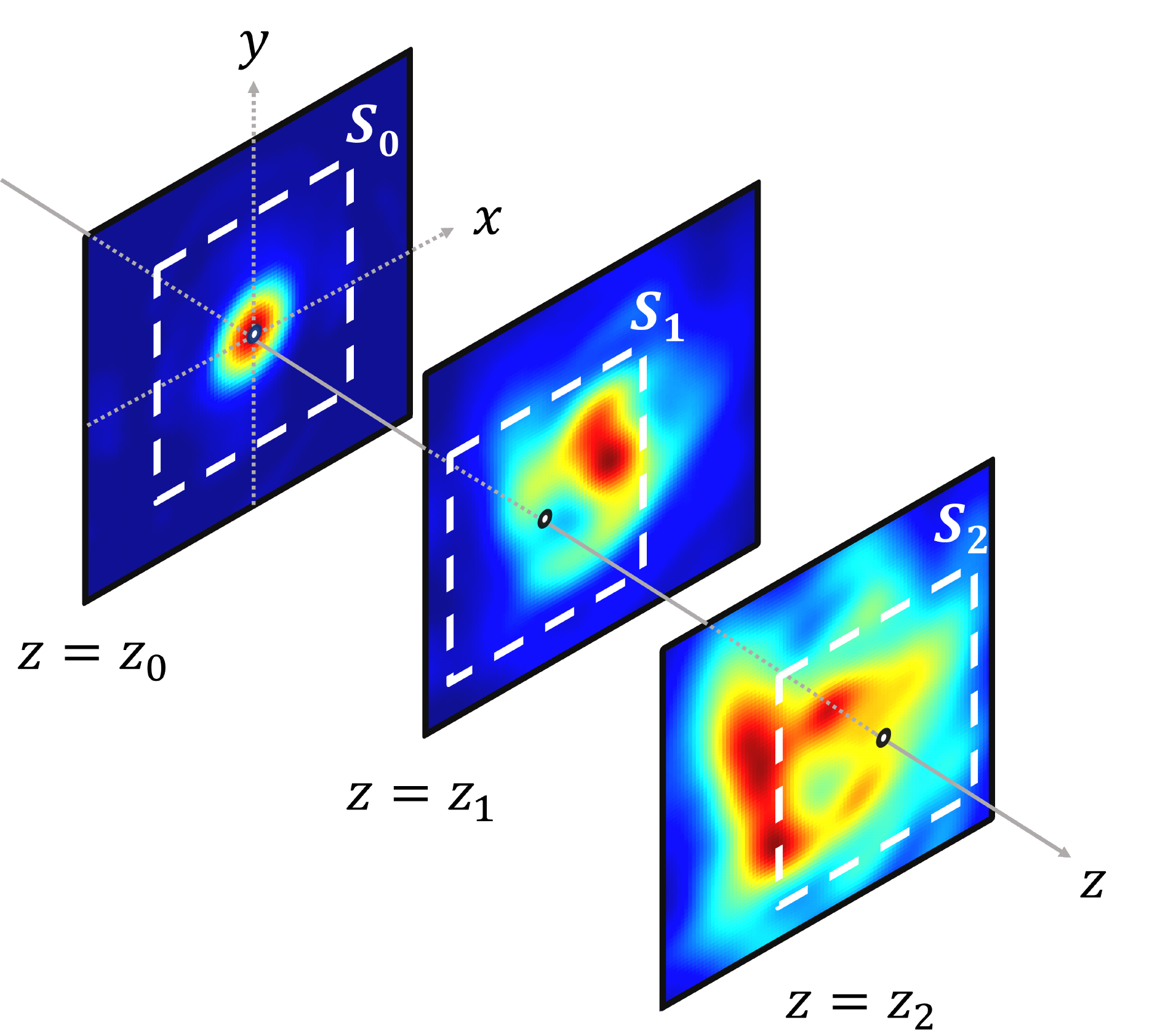}
\caption{Example of set of 3 fluence images $F_{exp}(x,y,z_k)$ and notations used for the GSA-MD calculation:
$z$ axis is the propagation axis originating from the center of energy of an image chosen as a reference (here $k=0$).
White dashed line are the search areas $S_k$ defined for the mode center tuning described in subsection 2 \ref{sec:centers_tuning}.
The mode centers in plane $k$, ($x_{0,k}$,$y_{0,k}$), are searched within $S_k$ and do not necessarily lie on the same $z$ axis. The fluence images come from different laser shots. The plane at $z=z_0$ is the focal plane. In this case it is the position of the first available measurement along the propagation axis, but in the general case the position $z=z_0$ may lie between the positions $z_k$ of other measurement planes. }
\label{fig:reference}
\end{figure}

\section{Overview of the field reconstruction method}

The proposed GSA-MD aims  to reconstruct the laser field of an electromagnetic wave propagating in the $z$ direction from experimentally obtained fluence images $F_{exp}(x,y,z_k)$, measured at different longitudinal distances $z_k$ from the focal plane and obtained from different shots of the same laser system, as illustrated in Fig. \ref{fig:reference}.

A laser pulse with carrier angular frequency $\omega_0$ and with negligible STC, propagating in the $z$ direction, can be described as a plane wave with transverse electric field $\mathcal{E}(x,y,z)$ and transverse complex envelope $\mathrm{E}(x,y,z)$ modulated by a temporal profile $T\left(t-\frac{z}{c}\right)$:
\begin{equation}\label{eq:envelope}
\mathcal{E}(x,y,z)=\mathrm{Re}\left\{\mathrm{E}(x,y,z)T\left(t-\frac{z}{c}\right)\mathrm{exp}\left[i\omega_0\left(t-\frac{z}{c}\right)\right]\right\},
\end{equation}
where $c$ is the velocity of light in vacuum.
Under the paraxial approximation, 
%and with the temporal profile assumed to be of the form $T\left(t-\frac{z}{c}\right)$, 
the laser field complex envelope can be decomposed as a sum of Hermite-Gauss (HG) modes:
\begin{equation}\label{eq:HGdecomposition}
\mathrm{E}(x,y,z) = \sum_{m,n}^{N_{m},N_{n}} C_{mn}HG_{mn}(x,x_0,y,y_0,z),
\end{equation}
\noindent where the modes $HG_{m,n}(x,x_0,y,y_0,z)$ are orthonormal and $N_{m}$ and $N_{n}$ are the number of modes in the $x$ and $y$ directions respectively for the HG modes expansion. The centers of the HG modes in the $x$ and $y$ directions are respectively $x_0$ and $y_0$. The values of these centers are not specified \textit{a priori}, and are part of the unknowns for the GSA-MD.

The HG modes of Eq. (\ref{eq:HGdecomposition}) are defined as \cite{Siegman86}:

\begin{align}\label{eq:HGmodes2}
HG_{m,n}(x, x_0, y, y_0, z) &= HG_m(x,x_0,z)\thinspace HG_n(y,y_0,z)\thinspace\exp\left[i\Phi(z)\right]\nonumber\\
HG_m(x,x_0,z)&=A_m \thinspace h_m\left[\sqrt{2}\frac{(x-x_0)}{w_{x}(z)}\right]\exp\left[-\frac{(x-x_0)^2}{w^2_{x}(z)}\right]\nonumber\\
&\times\exp\left[-ik_0\frac{(x-x_0)^2}{2R_{x}(z)}\right];\nonumber\\
HG_n(y, y_0, z)&=A_n \thinspace h_n\left[\sqrt{2}\frac{(y - y_0)}{w_{y}(z)}\right]\exp\left[-\frac{(y - y_0)^2}{w^2_{y}(z)}\right]\nonumber\\
&\times\exp\left[-ik_0\frac{(y - y_0)^2}{2R_{y}(z)}\right];\nonumber\\
\frac{w_{x}(z)}{w_{0,x}}&=\sqrt{1+\left(\frac{z}{Z_x}\right)^2};
\frac{w_{y}(z)}{w_{0,y}}=\sqrt{1+\left(\frac{z}{Z_y}\right)^2};\nonumber\\
A_m &= \left(w_{x}(z)2^{m - 1/2}m!\sqrt{\pi}\right)^{-1/2};\nonumber\\
A_n &= \left(w_{y}(z)2^{n - 1/2}n!\sqrt{\pi}\right)^{-1/2};\nonumber\\
R_{x}(z)&=z+\left(\frac{Z_x^2}{z}\right);R_{y}(z)=z+\left(\frac{Z_y^2}{z}\right);\nonumber\\
\Phi(z)&=\Phi_{x}(z)+\Phi_{y}(z);\nonumber\\
\Phi_{x}(z)&=\left(m+\frac{1}{2}\right)\tan^{-1}\left(\frac{z}{Z_x}\right);\nonumber\\
\Phi_{y}(z)&=\left(n+\frac{1}{2}\right)\tan^{-1}\left(\frac{z}{Z_y}\right),
\end{align}
where $h_k$ is the Hermite polynomial of order $k$.
The waists $w_{0x}=(2Z_x/k_0)^{1/2}$, $w_{0y}=(2Z_y/k_0)^{1/2}$ of the HG modes in the $x$, $y$ directions are chosen small enough to let the mode field reach negligible values at the borders of the measured images, and large enough to have Rayleigh lengths $Z_x$ and $Z_y$ which allow propagation up to the measurement planes. They may not be equal to the waists of a Gaussian fit of the fluences. The plane $z=0$ is chosen as the focal plane, i.e. where $w_{x}=w_{0,x}$ and $w_{y}=w_{0,y}$.
The uncertainty $\Delta_z$ on the focal plane position is taken into account in subsection 2 \ref{sec:field_reconstruction_algorithm}.

%It is important to highlight that the fluence images $F(x,y,z_k)$ may come from the average of different laser shots, which may have shot-to-shot wavefront and pointing fluctuations.
%Thus, for the purpose of field reconstruction using the data from $F_{exp}(x,y,z_k)$, the centers of the HG modes $x_{0,k}$, $y_{0,k}$ at different $z_k$ positions do not necessary lie on the same line (see Fig. \ref{fig:reference}).

The real and imaginary parts of the HG coefficients $C_{mn}$ are the unknowns.
Uncertainties in the laser fluence measurements arise from shot-to-shot fluctuations since transverse laser images taken at different positions with the same detector required different shots.
The quality of the field reconstruction depends on the reproducibility of the laser properties from shot to shot. Therefore, the field reconstruction consists in  fitting  fluence images to infer the corresponding laser field's amplitude and phase, taking into account shot-to-shot wavefront and pointing fluctuations. In the following, this process is referred to as the reconstruction of the laser field.

The measured fluence images are preprocessed as follows: first the background value is subtracted, then fluence values below a fixed threshold are put to zero, and each image is smoothed by pre-projecting it on a high number of HG modes assuming a phase uniformly equal to zero. The energy distribution centroids in $x$, $y$ are calculated for each position $z_k$. Then, each measured image is recentered on its centroid. Finally, the fluence of the measured images is divided by a fixed normalizing energy value $E_{norm}$.

The proposed GSA-MD aims at minimizing an error $\chi^2$ associated to the field reconstruction, defined as:

\textcolor{black}{
\begin{equation}\label{eq:error_chi}
\chi^2=\sum_{k=0}^{N_{images}-1}\dfrac{\sqrt{\sum_{i_x,\thinspace i_y}^{N_{pix_x},N_{pix_y}} (F_{exp}(x,y,z_k)-F_{fit}(x,y,z_k))^2}}{N_{images}\sum_{i_x,\thinspace i_y}^{N_{pix_x},N_{pix_y}} F_{exp}(x,y,z_k)},
\end{equation}
}

where $N_{pix_x},N_{pix_y}$ are the number of pixels of the image in the $x$ and $y$ directions, $F_{exp}$ and $F_{fit}$ are the measured and reconstructed fluences, $z_k$ are the positions of the $N_{images}$ measured images used for the reconstruction. $\chi^2$ in Eq. (\ref{eq:error_chi}) quantifies the error between the measured fluence data and the reconstructed fluence images. Although other error metrics can be chosen, without loss of generality it is assumed in the following that the chosen error metric is the $\chi^2$  in Eq. (\ref{eq:error_chi}). 

The evaluation of Eq. (\ref{eq:error_chi}) is computationally expensive in typical conditions of interest, for example using 3 images with $N_{pix_x}\times N_{pix_y}=1000\times1000$ pixels. Besides, the number of unknowns in Eq. (\ref{eq:HGdecomposition}), i.e. the real and imaginary parts of the reconstruction coefficients $C_{mn}$, is $2\times N_{m}\times N_{n}$, with typical values of $N_{m}=N_{n}$ of the order of 30, yielding 1800 unknowns. Furthermore, while the centers of the HG modes reconstruction of Eq. (\ref{eq:HGdecomposition}) in the plane $z_0$ can be fixed at the point of maximum fluence at $z=z_0$, the error of the reconstruction depends also on the chosen HG modes centers $(x_{0,k}, y_{0,k})$ in the other planes $z_k$. Thus, the choice of these centers must be optimized as well. If they are counted as additional degrees of freedom in the field reconstruction, the total number of unknowns is $2\times(N_{images})$ times larger. For the sake of comparison, it is worth noting that for a field reconstruction with a GSA, the number of unknowns (the phase values of each pixel) would be $N_{pix_x}\times N_{pix_y}$, i.e. $\mathrm{10^6}$ in the previous example. Therefore, in these conditions a direct minimization of $\chi^2$, optimizing at the same time the HG coefficients $C_{mn}$ and the HG centers $x_{0,k}$, $y_{0,k}$ would be too computationally expensive. 

The GSA-MD proposed in this article separates the search of the HG coefficients $C_{mn}$ for given values of the HG modes centers $(x_{0,k}, y_{0,k})$, and the search for the values of these centers that minimize the reconstruction error $\chi^2$. An additional advantage of this two-fold strategy is that the techniques used to address each of these two problems can be chosen independently.
For example basis functions different from the HG modes could in principle be used to find the expansion coefficients, without changing the technique used to optimize mode centers.

\begin{figure}[ht]
\centering
\includegraphics[width=\linewidth]{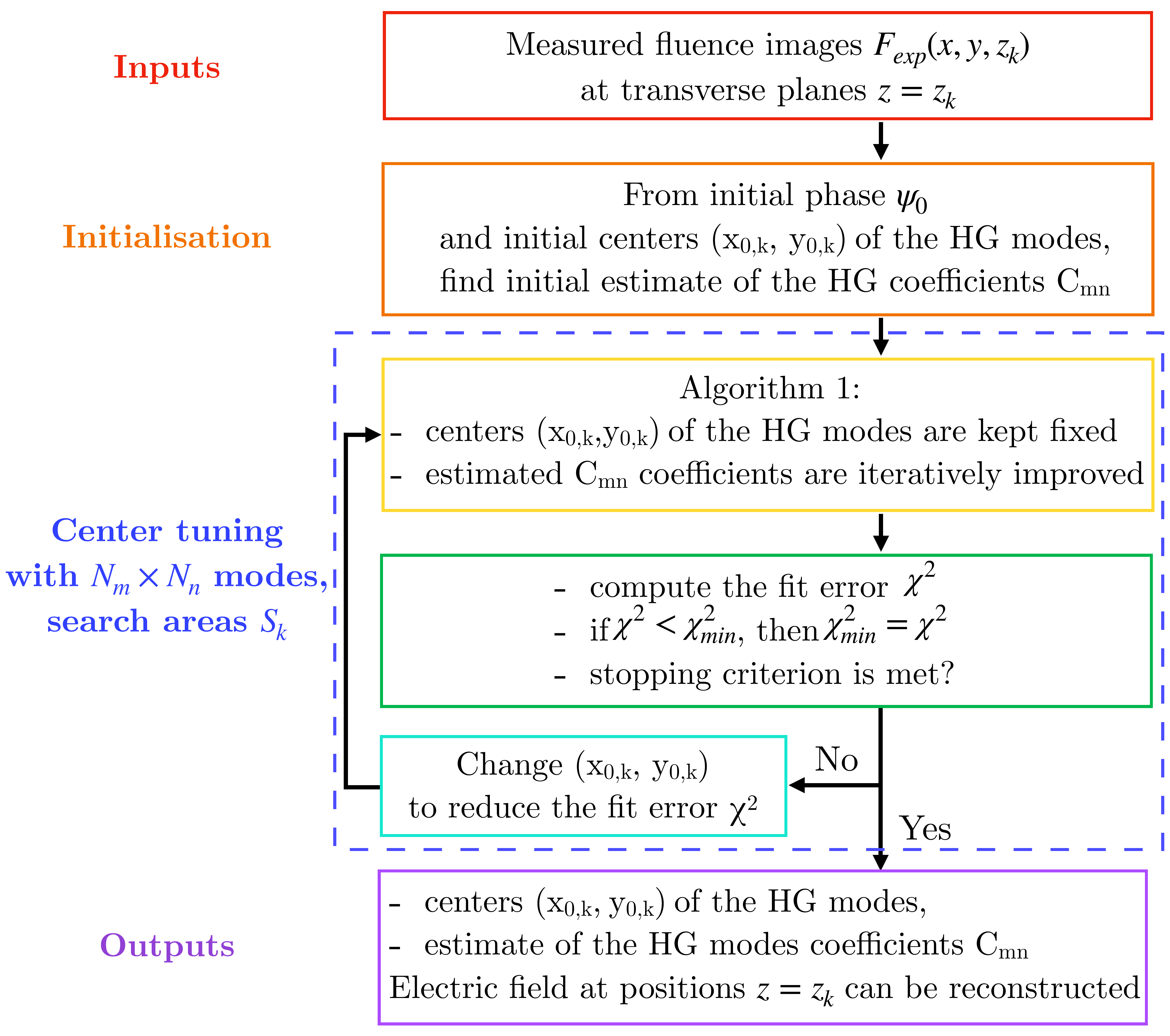}
\caption{Schematic overview of the proposed GSA-MD to reconstruct the laser field.
The yellow rectangle contains Algorithm \ref{alg:field_reconstruction}, detailed in Section 2\ref{sec:field_reconstruction_algorithm}. The tuning of the HG mode centers (blue dashed rectangle), performed to reduce the reconstruction error $\chi^2$, is described in Section 2\ref{sec:centers_tuning}.}
\label{fig:method_loop}
\end{figure}

This conceptual separation of the two mentioned problems is illustrated in Fig. \ref{fig:method_loop}, which gives an overview of the GSA-MD. The input of the GSA-MD is the fluence data $F_{exp}(x,y,z_k)$, measured in the transverse planes at position $z_k$. After preprocessing the fluence data, an initialization step is performed, which consists in finding an initial approximation of the HG coefficients $C_{mn}$ starting from an initial phase $\psi_0(x, x_{0,0}, y, y_{0,0})$ and an initial value for the HG modes centers $(x_{0,k}, y_{0,k})$.

Then, for fixed values of the HG modes centers $(x_{0,k},y_{0,k})$, the HG coefficients $C_{mn}$ estimates are improved iteratively. This update of the $C_{mn}$ coefficients is summarized in Algorithm \ref{alg:field_reconstruction} and detailed in the next section. The resulting reconstruction error $\chi^2$ in Eq. (\ref{eq:error_chi}) is then computed. Afterwards, the HG modes centers $(x_{0,k}, y_{0,k})$
can be changed in order to reduce the error $\chi^2$, and the $C_{mn}$ are updated using these new centers. If the new $\chi^2$ is lower than the minimum error $\chi^2_{min}$ found in this loop, the new $\chi^2$ substitutes the minimum error $\chi^2_{min}$.
A stopping criterion for this loop is chosen, e.g. reaching a maximum number of loop iterations or when the minimum error $\chi^2_{min}$ is reduced below a desired value. 

When the GSA-MD exits this loop, the resulting outputs will be values of the HG modes centers $(x_{0,k}, y_{0,k})$ and of the HG coefficients $C_{mn}$ that can be used to reconstruct the electric field at the planes $z_k$ using Eqs. (\ref{eq:HGdecomposition}), (\ref{eq:HGmodes2}). 

The next subsections  describe the update of the $C_{mn}$ coefficients (performed with fixed HG modes centers) and the search for the best choice of the HG mode centers.

%With square images ($N_{pix_x}=N_{pix_xy}=N_{pix}$) and equal number of modes in each direction ( $N_{m}$=$N_{n}$), as long as $N_{m}<N_{pix}/\sqrt{2}$ the number of unknowns involved in the field reconstruction with the algorithm in this article will be lower than the one with the GSA. With $N_{pix}=1000$, this condition corresponds to use less than $\approx700$ modes, which is much higher than the typical number of modes used with the presented algorithm.

\subsection{Calculation of the Hermite-Gauss modes coefficients}\label{sec:field_reconstruction_algorithm}

In this section an iterative algorithm is presented, to find the HG coefficients $C_{mn}$ of Eq. (\ref{eq:HGdecomposition}) that fit the laser transverse electric field, once the HG modes centers $(x_{0,k},y_{0,k})$ and waists $w_{0,x}$, $w_{0,y}$ are kept fixed, i.e. the algorithm in the yellow rectangle of Fig. \ref{fig:method_loop}.

%In many experimental campaigns with TW/PW class femtosecond lasers, only the fluence maps $F(x,y,z)$ of the laser pulse at specific distances $z_k$, with $k=0,1,...,N_{images}-1$ from the focal spot are known.

Assuming that no STC are present in the laser field, once the temporal profile $T(t-z/c)$ in Eq. (\ref{eq:envelope}) for the laser field is known (or a hypothesis on its shape is assumed), a linear relation between the experimentally measured fluence $F_{exp}(x,y,z)$ and local intensity $I(x,y,z)$ can be easily obtained, i.e. $F_{exp}(x,y,z)=I(x,y,z)\cdot \tau$, where $\tau$ is a characteristic duration of the laser pulse and the local intensity is defined as $I(x,y,z)=\frac{c\varepsilon_0}{2}|\mathrm{E}(x,y,z)|^2$.

A complex envelope $\mathrm{E}$  of the transverse electric field at position $z$ can thus be defined from a phase map $\psi(x,y)$ and an experimental fluence map $F_{exp}(x,y,z)$:
\begin{equation}\label{eq:e_reconstruction}
\mathrm{E}(x,y,z) =\sqrt{\frac{2}{c\tau\varepsilon_0}F_{exp}(x,y,z_0)}\thinspace\exp{[i\psi(x,y)]},
\end{equation}
where $\varepsilon_0$ is the vacuum permittivity. 

As in the classic GSA, this operation is performed at the available measurement planes combining the intensity $I$, expressed in this article in terms of measured fluence $F_{exp}$ after assuming a temporal profile, and the estimated phase map $\psi(x,y)$.

Using this definition, the calculation of the HG coefficients $C_{mn}$ for the field reconstruction is summarized by the pseudocode in Algorithm \ref{alg:field_reconstruction}, which is described in the following.

\begin{algorithm}[ht]
\caption{Algorithm to find the coefficients $C_{mn}$ of the Hermite-Gauss modes $HG_{mn}$ from $N_{images}$ experimental fluence images $F_{exp}$ measured at planes $z_k$, with $k=0,...,N_{images}-1$. The HG modes centers $(x_{0,k}, y_{0,k})$ are set at the start of the algorithm and kept fixed. Steps 6-9 are repeated for each of the mode indices $m$, $n$. This algorithm corresponds to the yellow rectangle of Fig. \ref{fig:method_loop}
.}\label{alg:field_reconstruction}
    \begin{algorithmic}
        \Procedure{Field reconstruction}{}\\
            \State 1) Find an initial estimate of $C_{mn}$;
            \For{($iter=0$;\quad$iter<N_{iter}$;\quad$iter++$)}\\
            
            \For{($k=0$;\quad$k<N_{images}$;\quad$k++$)}\\
            \State 2) $\mathrm{E}(x,y,z_k)=$
            \State $=\sum_{m,n} C_{mn}HG_{mn}(x,x_{0,k},y,y_{0,k},z_k)$;
            \State 3) $\psi(x,y)=\arg{\left[\mathrm{E} (x,y,z_k)\right]}$;
            \State 4) $\mathrm{E_{new}}(x,y,z_k)=$ 
            \State $=\sqrt{\frac{2}{c\tau\varepsilon_0}F_{exp}(x,y,z_k)}\exp{[i\psi(x,y)]}$;
            \State 5) $\delta(x,y,z_k)=$
            \State $=\dfrac{\sqrt{\frac{2}{c\tau\varepsilon_0}F_{exp}(x,y,z_k)} - |\mathrm{E}(x,y,z_k)|}{\mathrm{max}\left[\sqrt{\frac{2}{c\tau\varepsilon_0}F_{exp}(x,y,z_k)}\right]}$;
            \State $\mathrm{E_{new}}(x,y,z_k)= \mathrm{E_{new}}(x,y,z_k)\thinspace\exp{[\delta(x,y,z_k)]}$;
            \State 6) $C_{mn,k}=\mathrm{Proj}[\mathrm{E_{new}}(x,y,z_k)$\\
            \hspace{3.6cm}$,HG_{mn}(x,x_{0,k},y,y_{0,k},z_k)]$;
            \State 7) $C_{mn,k}=C_{mn,k}\thinspace \sqrt{\dfrac{F_{tot}}{\sum_k|C_{mn,k}|^2}}$;
            \State 8) $C_{mn}=\frac{1}{2}\left(C_{mn}+C_{mn,k}\right)$;
            \State 9) $C_{mn}=C_{mn}\thinspace \sqrt{\dfrac{F_{tot}}{\sum_k |C_{mn}|^2}}$;
        \EndFor
        \If{($iter\%5==0$) and ($iter\geq5$)}\\
            \State 10) $\chi^2_{grad}$ = $\dfrac{\chi^2(iter) - \chi^2(iter-5)}{\chi^2(iter - 5)}$;\\
            \If{($\chi^2_{grad} < 0.02$)}\\
                \State 
                $iter_{break}=iter$;\\
                \State 
                \textbf{break};
            \EndIf
        \EndIf
    \EndFor
    \EndProcedure
\end{algorithmic}
\end{algorithm}

First, an initial estimate of the coefficients is computed (step 1). This first estimate can be obtained from a first projection of $\sqrt{\frac{2}{c\tau\varepsilon_0}F_{exp}(x,y,z_0)}\exp\left[\psi_0(x, x_{0,0}, y, y_{0,0})\right]$ over the HG modes with an initial choice of the modes centers $x_{0,k}$, $y_{0,k}$ and initial phase $\psi_0(x, x_{0,0}, y, y_{0,0})$.

For the results presented in this article, to improve the convergence of the field reconstruction, an initial quadratic phase $\psi_0(x, x_{0,0}, y, y_{0,0})$ was used (similar to the initial phase proposed in \cite{pang2017non}):
\begin{equation}\label{eq:quadratic_phase}
    \psi_0(x, x_{0,0}, y, y_{0,0}) = k_0 \dfrac{(x-x_{0,0})^2+(y-y_{0,0})^2}{2 \Delta_z \left[1+\left(\frac{k_0}{2}\frac{ w_0^2}{\Delta_z}\right)^2\right]},
\end{equation}
where $w_{0,Gauss}$ is the estimated waist of a Gaussian fit of the measured fluence map $F_{exp}(x,y,z_0)$. This initial phase represents the phase of a Gaussian beam with waist $w_0$ and carrier frequency $\omega_0$, at a distance $\Delta_z$, which is the uncertainty on the focal plane $z$ position. 

After this initialization, at each iteration $iter$ of the algorithm, the estimated expansion of $\mathrm{E}(x,y,z_k)$ in HG modes $HG_{mn}(x,x_{0,k},y,y_{0,k},z_k)$ is computed at each position from $z_0$ to $z_{N_{images}-1}$, using the known expressions of the HG modes \cite{Siegman86} (Eq. (\ref{eq:HGmodes2})) and the estimated coefficients $C_{mn}$, using Eq. (\ref{eq:HGdecomposition}) (step 2).
The phase map $\psi(x,y)$ is then found as $\arg\left[\mathrm{E}(x,y,z_k)\right]$ (step 3). 

In step 4, an updated value of the complex electric field $\mathrm{E_{new}}$ can be estimated using the measured fluence $F(x,y,z_k)$ and the phase $\psi(x,y)$, using Eq. (\ref{eq:e_reconstruction}).

The exponent $\delta(x,y,z_k)$ of an exponential correction factor $\exp\left[\delta(x,y,z_k)\right]$ is calculated on each point of the grid. The resulting correction factor is equal to one at the points where the measured and reconstructed field amplitude are equal and its value is higher where the two amplitudes differ. The field $\mathrm{E_{new}}$ is multiplied by this correction factor (step 5). In \cite{wu2021adaptive} it has been shown that this correction improves the convergence of a GSA as well as the signal to noise ratio of its reconstruction. 

The projection of the corrected $\mathrm{E_{new}}$ on the HG modes at $z_k$ gives a new estimate $C_{mn,k}$ for the HG coefficients (step 6), which is combined with the previous estimate of $C_{mn}$  (step 8).

The projection of a function $f(x,y,z_k)$ on the HG modes at $z_k$ mentioned in step 6 is defined as:
\begin{eqnarray}\label{projectionHG}
\mathrm{Proj}[f(x,y,z_k),HG_{mn}(x,x_{0,k},y,y_{0,k},z_k)] =\nonumber \\
=\int_{-L_x/2}^{L_x/2}\int_{-L_y/2}^{L_y/2} f(x,y,z_k) HG_{mn}^*(x,x_{0,k},y,y_{0,k},z_k) dx\thinspace dy,
\end{eqnarray}
where $\left( L_x,L_y \right)$ are the data  grid length along each axis.

Normalizations are performed on the estimated coefficients in the intermediate steps 7 and 9 to ensure that the total fluence $F_{tot}$ remains constant.

Steps 6-9 are repeated for each index $m$, $n$ of the modes used in the field reconstruction. 

\textcolor{black}{In step 10), starting from $iter=0$ and every 5 iterations, the $\chi^2$ error is evaluated. If at a given iteration $iter$, the error gradient $\chi^2_{grad}$ is less than 2\%, then Algorithm 1 loop is stopped and the last iteration is recorded as $iter_{break}$.}
%After the last iteration $iter =N_{iter}-1$, the coefficients $C_{mn}$ are updated again for the plane $k=0$.

It is worth noting that the most computationally expensive operations of the algorithm are step 2, i.e. the reconstruction of the field with propagated HG modes, and step 6, i.e. the projection over the HG modes. 
This consideration highlights an advantage of the GSA-MD compared to the classic GSA: these two steps can be easily parallelized, since the treatment of each mode can be performed in parallel, with step 2 only requiring a final summation of the contribution of each mode.

The use of mode expansion yields two additional advantages compared to a classic GSA. First, in principle another set of basis function can be used instead of the HG modes, depending on the application. Second, the number of modes can be chosen in order to find the desired compromise between reconstruction accuracy and computation time. This latter flexibility will be illustrated in section 2\ref{sec:centers_tuning}.

As stated at the start of this subsection, in the algorithm it was assumed that the HG modes centers $(x_{0,k},y_{0,k})$ were set.
The next subsection describes how the choice of these centers can be improved to reduce the reconstruction error.

% \begin{itemize}
% \item Find an initial estimate of the coefficients $C_{m,n}(x,y)$ projecting the intensity corresponding to the fluence $F(x,y,z_0)$ on the $\mathrm{HG_{m,n}}(x,y,z_0)$ modes at $z_0$. In the following we denote the projection of a function $f(x,y,z)$ on the HG modes with the notation $\mathrm{Proj}$:
% \begin{equation}
% C_{m,n}(x,y)=\mathrm{Proj}[f(x,y,z),\mathrm{HG_{n,p}}(x,y,z)].\\
% \end{equation}
% \item For $iter=0 \rightarrow N_{\mathrm{iter}}$ and for $k=0 \rightarrow k_{\mathrm{max}}$:
% \begin{itemize}
% \item define the propagated field as 
% \begin{equation}
% \mathrm{E}_{GSA,z_k} (x,y,z_k)=\sum_{m,n} C_{m,n}(x,y)\cdot \mathrm{HG_{m,n}}(x,y,z_k);
% \end{equation}
% \item find the phase map $\psi$ as:
% \begin{equation}
% \psi(x,y)=\arg{(\mathrm{E}_{GSA,z_k} (x,y,z_k))};
% \end{equation}
% \item combine the measured fluence $F(x,y,z_k)$ with the phase $\psi(x,y)$ to find the function $\mathrm{E'}_{GSA,z_k}$ using Eq. \ref{eq:e_reconstruction}:
% \begin{equation}
% \mathrm{E'}_{GSA,z_k} (x,y,z_k)= \mathrm{E}[F(x,y,z_k),\psi(x,y)];
% \end{equation}
% \item combine the previous estimate of the HG coefficients with those obtained from the projection of $\mathrm{E'}_{GSA,z_k}$ on the HG modes at $z_k$:
% \begin{eqnarray}
% C_{m,n}(x,y)=(1 - \alpha)\cdot C_{m,n}(x,y)+\nonumber\\ 
% +\alpha\cdot\mathrm{Proj}[\mathrm{E'}_{GSA,z_k},\mathrm{HG_{n,p}}(x,y,z_k)];
% \end{eqnarray}

% \end{itemize}

% \end{itemize}

\subsection{Tuning the centers of the Hermite-Gauss modes}\label{sec:centers_tuning}

The error of the reconstruction algorithm of the section 2\ref{sec:field_reconstruction_algorithm} is sensitive to the choice of the HG mode centers $(x_{0,k}, y_{0,k})$. Thus, as shown in Fig. \ref{fig:method_loop}, the field reconstruction in Algorithm \ref{alg:field_reconstruction} can be repeated with different $(x_{0,k},y_{0,k})$ chosen within a search area $S_k$ at each plane $z_k$ (see Fig. \ref{fig:reference}) in order find their values which minimize (or at least reduce) the reconstruction error.

The separation of the HG coefficient estimation in Algorithm~\ref{alg:field_reconstruction} from this tuning of the HG mode centers $(x_{0,k},y_{0,k})$ allows to choose among many optimization algorithms to minimize the error $\chi^2$. For example, Bayesian optimization \cite{Frazier2018} was used for the results presented in section \ref{sec:results}. In the following, this general minimization process is referred to as the center tuning, which is stopped when a chosen criterion is met, e.g. when a certain target value of $\chi^2$ is reached, or when a total number of iterations $N_{tuning}$ is completed.

In general the quality of the field reconstruction is sensitive to the combination of the main parameters of the GSA-MD, namely $N_{m}$, $N_{n}$, $N_{iter}$, $N_{tuning}$ and the size of the projection grid. Increasing these parameters yields a longer computing time for the field reconstruction in Algorithm~\ref{alg:field_reconstruction} and the center tuning. They can be set depending on the quality of the available fluence data (e.g. degree of asymmetry) in order to find a compromise between reconstruction accuracy and computing time required by the minimization of the error $\chi^2$.

As previously mentioned, decomposing the field with HG modes introduces a flexibility in the choice of the number of modes $N_{m}$, $N_{n}$ (along the $x$ and $y$ directions respectively) used for the reconstruction in Algorithm \ref{alg:field_reconstruction}. This flexibility can be used to speed-up the center tuning, as explained in the next section.

\section{Results}\label{sec:results}

%\hl{For the results presented in the next section, the HG modes origin tuning is performed through Bayesian Optimization of the function $\chi^2$. 
%by Gaussian process regression and picked within a reduced search area $S=$\{[$x_{min}$; $x_{max}$], [$y_{min}$; $y_{max}$]\}.
%In both phases, the optimizer is ran through a cycle of $N_{tuning,EG,RS}$ iterations, each one consisting of $N_{iter}$ loops of the Reconstruction Algorithm. Since time consuming steps of the algorithm, namely the Projection and Propagation, are being parallelized with Numba, the Bayesian Optimization is executed in threads over $n_{threads}$ concurrent workers which constant value is set prior to the tuning. The optimizer is called in parallel over $N_{calls} = N_{tuning}\/\/n_{threads} + N_{tuning}\%n_{threads}$. At each called iteration $n_{call}$, if $n_{call} \leq n_{threads} \times \left(N_{tuning}\/\/n_{threads}\right)$, then a number of $n_{threads}$ centers set {$x_k$, $y_k$, $k = 1...N_{images}$} are drawn based on the optimizer prior data at the moment of the call. WhenThe origin at the reference plane, which is also the focal plane, is fixed to $x_0=y_0=0$.\\
%The optimizer uses an implementation of the standard linear regression model with Gaussian noise introduced in Algorithm 2.1 of \cite{williams2006gaussian}, with a default "1.0 * RBF(1.0)" kernel. 
%After all
%and $N_{tuning}$. 
%$N_{tuning,RS}$
 %, with (x=0, y=0) referring to the centroid of the focal plane measured energy distribution.}

In this section the results of the GSA-MD, applied on laser data collected at the LLC (peak power in the data 23 TW, pulse duration 38 fs), and on the Apollon laser system in the commissioning phase (peak power in the data 400 TW, pulse duration 25 fs), are presented.

\textcolor{black}{For both campaigns, fluence measurements were performed using a CCD camera equipped with a microscope objective, which was translated along the laser axis in the focal volume in vacuum. For these measurements, the laser beam was fully amplified to nominal energy, then attenuated by several reflections from glass surfaces before compression, in order to characterize the quality of the high intensity beam. 
At every position of the camera along the laser axis, multiple measurements were made in order to evaluate the shot-to-shot fluctuations of the laser.}

The pointing stability for both data-sets is characterised by the shot-to-shot fluctuations of the fluence centroids normalized by the estimated laser waist $\delta \overline{x}/w_{0,Gauss}$, $\delta \overline{y}/w_{0,Gauss}$, where $w_{0,Gauss}$ is the estimated Gaussian fit's waist. For the LLC data-set,  $\delta \overline{x}/w_{0,Gauss}=16$ $\%$ and $\delta \overline{y}/w_{0,Gauss}=8$ $\%$, with $w_{0,Gauss}=15$ $\mu$m. For the Apollon data-set, the shot-to-shot pointing instability is higher: $\delta \overline{x}/w_{0,Gauss}=100$ $\%$ and $\delta \overline{y}/w_{0,Gauss}=40$ $\%$, with $w_{0,Gauss}=16$ $\mu$m. It will be shown that the GSA-MD can reconstruct the laser field from the fluence data of both these two different laser systems.

\begin{figure}[ht]
\centering
\includegraphics[width=\linewidth]{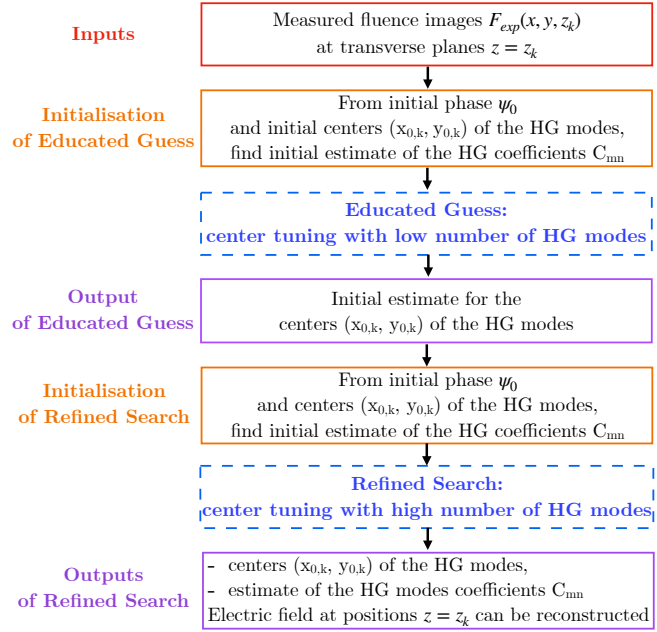}
\caption{Schematic description of the tuning of HG mode centers that was used to reduce the minimum field reconstruction error $\chi^2_{min}$ for the LLC and Apollon data-sets.}
\label{fig:tuning}
\end{figure}

Figure \ref{fig:tuning} describes the procedure used to obtain the results presented in this section, for the LLC and Apollon data-sets. This procedure exploits the GSA-MD's flexibility in choosing the number of modes for the field reconstruction.
Performing the the center tuning introduced in Fig. \ref{fig:method_loop} with a high number of HG modes would have been computationally expensive. Thus, the center tuning has been separated in two successive phases (blue dashed rectangles of Fig. \ref{fig:tuning}) that share the same Algorithm \ref{alg:field_reconstruction} and minimization method for the error $\chi^2$ (Bayesian Optimization in this case), but with a different number of HG modes $N_{m}$, $N_{n}$.

The first phase, referred to as the Educated Guess (EG), consists of a center tuning with $N_{tuning,EG}$ iterations, each using $N_{m,EG}$ and $N_{n,EG}$ modes set low enough to quickly execute Algorithm \ref{alg:field_reconstruction}. This EG phase can be initialized setting  $x_{0,k}=y_{0,k}=0$ as initial centers and Eq. \ref{eq:quadratic_phase} as initial phase $\psi_0(x, 0, y, 0)$.

This EG phase yields an initial estimate of the HG centers $x_{0,k}$ and $y_{0,k}$.
The phase $\psi_0(x, x_{0,0}, y, y_{0,0})$ of Eq. (\ref{eq:quadratic_phase}) is reinitialized with the optimized centers ($x_{0,0}$, $y_{0,0}$) tuned in the EG.
Using these centers, the phase and $F_{exp}(x,y,z_0)$, a projection of $\sqrt{\frac{2}{c\tau\varepsilon_0}F_{exp}(x,y,z_0)}\exp\left[i\psi_0(x, x_{0,0}, y, y_{0,0})\right]$ over the HG modes yields a more accurate estimate of the $C_{mn}$ coefficients, even with a different number of modes. 
This estimate is used to initialize a second center tuning phase, called Refined Search (RS), which is performed with a higher number of HG modes $N_{m,RS}$ and $N_{tuning,RS}$ center tuning iterations, using a narrower search area for the HG centers.

For the results with the LLC data-set, in Eq. (\ref{eq:quadratic_phase}), $w_{0}=15$ $\mu$m and $\Delta_z=0.25$ mm. For the Apollon data-set, $w_{0}=16$ $\mu$m and $\Delta_z=0.3$ mm. For both data-sets, $w_{0,x}=w_{0,y}=20$ $\mu$m has been used for the HG modes waists.

The implementation of the GSA-MD used for this article is written in Python. The most time consuming steps of Algorithm \ref{alg:field_reconstruction}, steps 2) and 6), are compiled and parallelized with Numba. 

To obtain the presented results, the HG mode center tuning in both EG and RS phases was performed through Bayesian Optimization \cite{Frazier2018} of the function $\chi^2$ defined in Eq. (\ref{eq:error_chi}). 
At each iteration of the Bayesian Optimization, multiple values of the HG centers are chosen in parallel to execute Algorithm \ref{alg:field_reconstruction} and compute the corresponding values of $\chi^2$. Each parallel execution of Algorithm \ref{alg:field_reconstruction}, corresponding to different values of the HG centers, is distributed between the available computing threads. 
In the Bayesian Optimization algorithm, these new values of the HG centers are chosen within the search areas $S_{EG}$, and $S_{RS}$, for the Educated Guess and Refined Search, respectively. Each evaluation of $\chi^2$ corresponding to different values of the HG centers is used by the Bayesian Optimization algorithm to build a surrogate model for the function $\chi^2$. The probability distribution of possible $\chi^2$ values is modeled by a Gaussian Process with mean and standard deviation. The covariance matrix of the process, or kernel, defines the correlation between the evaluated points $\chi^2$ score and the estimated values for non-evaluated points. The minimum error $\chi^2_{min}$ is updated each time a new minimum for the error $\chi^2$ is found during the iterations of the error minimization process. 

In both EG and RS phases, the Bayesian Optimization uses an implementation of the standard linear regression model with Gaussian noise introduced in Algorithm 2.1 of \cite{williams2006gaussian}. The "1.0 * RBF(1.0)" kernel, present in the Python library $\mathrm{scikit-optimize}$ \cite{head2020scikit}, was used, with RBF being the radial basis function kernel. To choose the next candidate centers to evaluate, an acquisition function is used, which calculates the point with the optimum combination of the mean and uncertainty values from the Gaussian process via a combination of the Expected Improvement, Negative \textcolor{black}{Probability} of Improvement and Lower Confidence Bound acquisition functions described in \cite{shahriari2015taking}. Based on a scoring value of these functions, one of the proposed centers is chosen for the evaluation. The Bayesian Optimization is initiated with $skopt.Optimizer$, $\xi=0$, which skews heavily the Expected Improvement towards exploitation of previous evaluated points. The other parameters are fixed to their default values in the $\mathrm{scikit-optimize}$ library. 
Table \ref{tab:parameters} summarizes the parameters of the two data-sets and the parameters used for the reconstruction, described also in the following subsections. The results of the GSA-MD applied on the two data-sets will be presented.

\subsection{Field reconstruction for the LLC data-set}
With the LLC system, the average energy per shot collected in 2021 for the data used in this article is 872 mJ, for an average laser pulse duration of 38 fs, which represents a peak power $P_0=23$ TW. The central wavelength is $\lambda_0=0.8$ $\mu$m, and the waist of a Gaussian fit of the data measured in the focal plane is estimated at $w_{0,Gauss}=15$ $\mu$m, which sets the Rayleigh length of the Gaussian fit to $z_R\simeq0.9$ mm.

\begin{table}[ht]
\small
\centering
\caption{Data-sets and reconstruction parameters: carrier wavelength $\lambda_{0}$, peak power $P_0$, mean energy per laser shot, shot-to-shot relative pulse centroid position fluctuations $\delta\bar{x}/w_{0,Gauss}$ and $\delta\bar{y}/w_{0,Gauss}$, position $z$ of the fluence measurement planes ($z=0$ is the focal plane), number of pixels in the fluence images, pixel size, estimated Gaussian fit's waist $w_{0,Gauss}$, uncertainty of the focal plane position $\Delta_z$, waists $w_{0,x}=w_{0,y}$ for the HG modes, number of modes $N_m$ and $N_n$ in the $x$ and $y$ direction for the EG (RS) phase $N_{m,EG}$, $N_{n,EG}$ ($N_{m,RS}$, $N_{n,RS})$, search area $S_{EG}$ ($S_{RS}$) for the centers of the EG (RS) phase, number of iterations $N_{iter}$ for Algorithm \ref{alg:field_reconstruction}, number of center tuning iterations for the EG (RS) phase $N_{tuning,EG}$ ($N_{tuning,RS}$), computing time for the EG and RS phases.  }
\begin{tabular}{ccc}
\hline
Parameter & LLC data-set & Apollon data-set \\
\hline
$\lambda_{0}$ & 0.8 $\mu$m & 0.8 $\mu$m \\
Peak power $P_0$ & 23 TW & 400 TW \\
Mean energy/shot & 0.872 J &  4.8 J \\
$\delta\bar{x}/w_{0,Gauss}$, $\delta\bar{y}/w_{0,Gauss}$ & $16$ $\%$, $8$ $\%$ &  $100$ $\%$, $40$ $\%$ \\
$z$ & $[0,0.5,1,1.5]$ mm & $[0,-1.8,1.2]$ mm \\
$N_{pix_x}\times N_{pix_y}$ & 351$\times$351 & 301$\times$301 \\
Pixel size & 1.13 $\mu$m & 0.85 $\mu$m \\
Estimated $w_{0,Gauss}$ & 15 $\mu$m & 16 $\mu$m \\
$\Delta_z$ & 0.25 mm & 0.3 mm \\
$w_{0,x}=w_{0,y}$ & 20 $\mu$m & 20 $\mu$m \\
$N_{m,EG}$, $N_{n,EG}$ & 10, 10 & 10,10 \\
$N_{m,RS}$, $N_{n,RS}$ & 30, 30 & 40,40 \\
$S_{EG}$ & 20 $\mu$m $\times$ 20 $\mu$m & 100 $\mu$m $\times$ 100 $\mu$m \\
$S_{RS}$ & 10 $\mu$m $\times$ 10 $\mu$m & 20 $\mu$m $\times$ 20 $\mu$m \\
$N_{iter}$ & 50 & 50 \\
$N_{tuning,EG}$, $N_{tuning,RS}$ & 300, 300 & 300, 300\\
Computing time, EG & \textcolor{black}{19 minutes} & \textcolor{black}{18 minutes} \\
Computing time, RS &\textcolor{black}{42 minutes} & \textcolor{black}{57 minutes} \\
\hline
\end{tabular}
  \label{tab:parameters}
\end{table}

\textcolor{black}{The LLC data-set used for the algorithm is a set of 4 transverse fluence profiles $F_{exp}(x,y,z_k)$ at $z_{0,1,2,3}=0$, $500$, $1000$ and $1500$ $\mu$m. For a given position $z_k$, the fluence profile $F_{exp}(x,y,z_k)$ is randomly selected among 15 individual shot measurements for $k\neq2$ and 17 shots for $k = 2$.}

For each individual shot, the average background over a $100\times100$ pixels region far from the transverse focal spot energy has been subtracted. Then, for each averaged image, the fluence has been filtered setting values below $1$ $\%$ of the absolute maximum to zero. Each measured distribution has then been smoothed by projecting them onto HG modes with $N_{m}=N_{n}=40$.
The projecting box over which the HG modes are fitted is a square grid of $351\times351$ pixels ($397$ $\mu$m $\times397$ $\mu$m) centered on the centroid of the fluence map in the focal plane ($z=z_0$). The size of the box is determined to ensure that the HG modes, whose characteristic transverse extension scales with $w_{0,x}\sqrt{m}$, $w_{0,y}\sqrt{n}$ in the transverse directions, decay to 0 before reaching the grid boundaries in the plane further from focus.

% The field reconstruction has been performed with numerical waists $w_{0x}=w_{0y}=20$ $\mu$m, $N_{iter}=100$.

For the Educated Guess, $N_{tuning,EG}=300$, $N_{m,EG}=N_{n,EG}=10$ and a search area $S_{EG}=($ 20 $\mu$m$\times$ 20 $\mu$m$)$, centered around the centroid of the fluence distribution at $z=z_0$ was chosen.

For the Refined Search, $N_{tuning,RS}=300$, $N_{m,RS}=N_{n,RS}=30$ and a search area $S_{RS}=($ 10 $\mu$m$\times$ 10 $\mu$m$)$ centered around the calibrated centers found by the Educated Guess were chosen.

\begin{figure}[ht!]
    \centering
    \includegraphics[width=0.48\textwidth]{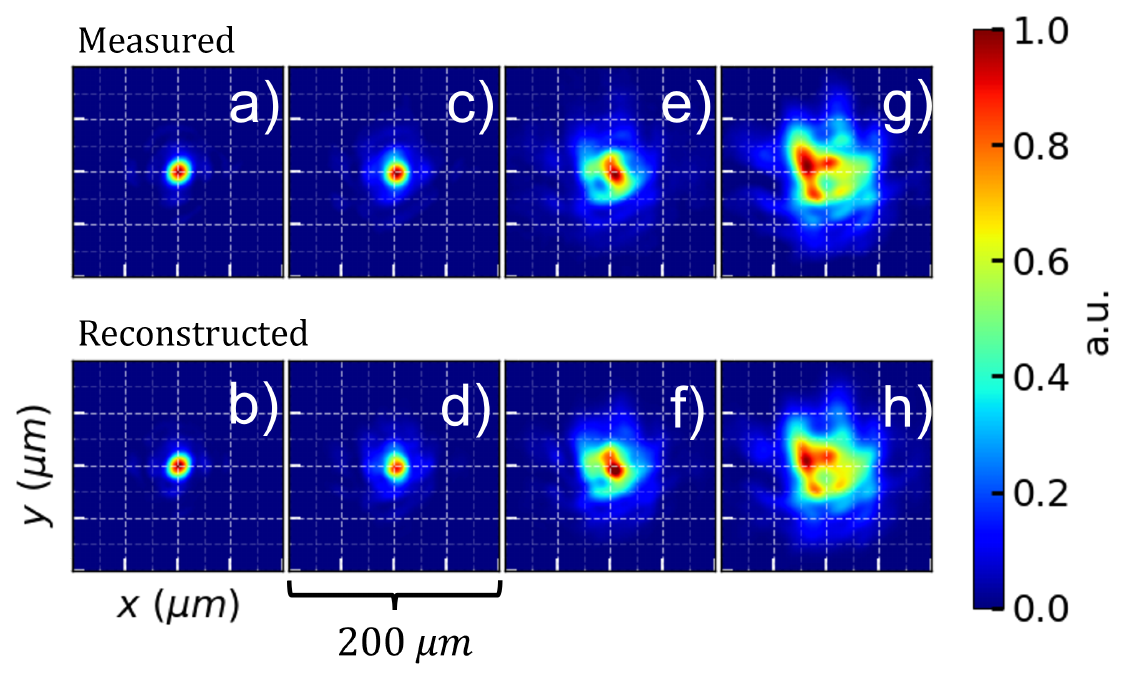}
    \caption{Measured fluence distribution of the LLC data-set (upper row) and corresponding reconstructed distributions after center tuning (lower row).  From left to right, the positions of the image planes along the propagation axis are :
    a, b) $z_0=0$ $\mu$m; c, d) $z_1=500$ $\mu$m ($0.6\thinspace z_R$); e, f) $z_2=1000$ $\mu$m ($1.1\thinspace z_R$); g, h) $z_3=1500$ $\mu$m ($1.7\thinspace z_R$).
    For each position $z_k$, the fluence has been normalized to the maximum of the corresponding measured fluence. }
    \label{fig:HG_fit}
\end{figure}

\begin{figure}[ht!]
    \centering
    \includegraphics[width=0.48\textwidth]{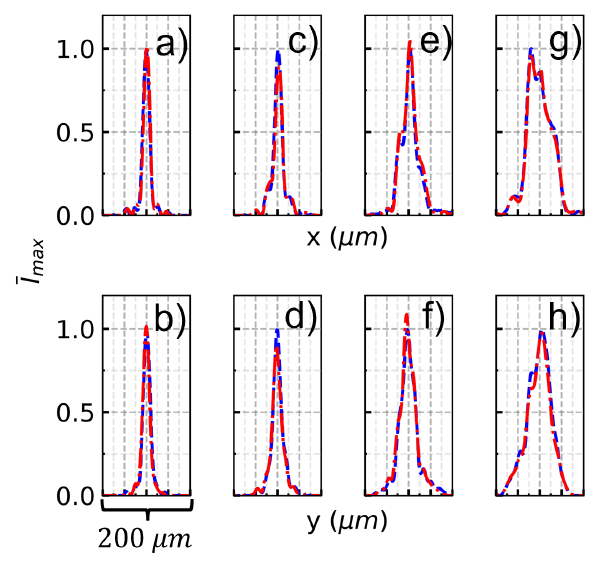}
    \caption{Fluence profiles along the $x$ (upper row) and $y$ (lower row) directions, averaged over 3 pixels (3.39 $\mu$m) centered around the measured fluence maximum's position in $y$ and $x$. Each profile has been normalized to the measured fluence maximum at $z_k$. For a given position $z_k$, the blue dashed line is the measured fluence from the LLC data-set and the red dashed line is the reconstructed fluence profile. From left to right, relative positions to the focal plane are :
    a, b) $z_0=0$ $\mu$m; c, d) $z_1=500$ $\mu$m ($0.6\thinspace z_R$); e, f) $z_2=1000$ $\mu$m ($1.1\thinspace z_R$); g, h) $z_3=1500$ $\mu$m ($1.7 \thinspace z_R$)}
    \label{fig:itg_fit}
\end{figure}

The final results of the GSA-MD calculation for the LLC data-set are shown in Figs. \ref{fig:HG_fit} and \ref{fig:itg_fit}.

Figure \ref{fig:HG_fit} shows the measured fluence images and the reconstructed fluence distributions at four positions along the propagation axis. Comparison of the images shows that the main features of the LLC data-set are well reconstructed by the GSA-MD calculation, in particular the asymmetries of the distribution at $z_2=1.1\thinspace z_R$ [Figs. \ref{fig:HG_fit} e), f)] and $z_3=1.7\thinspace z_R$ [Figs. \ref{fig:HG_fit} g), h)].

In Figure \ref{fig:itg_fit} the measured fluences in the $z_k$ planes and the corresponding reconstructed fluences are compared on 1D plots, for the data shown in Fig. \ref{fig:HG_fit}.
For each $z_k$ position, the fluence is plotted along the axis $x$ (top panel) and axis $y$ (bottom panel) directions, where the maximum measured fluence lie. Each line plot in the $x$ (resp. $y$) direction is an average over 3 pixels in the $y$ (resp. $x$) direction.
The maximum relative differences on the measured fluence's amplitude in x and y are \textcolor{black}{$2.4$}\% at $z_0$, \textcolor{black}{$10$}\% at $z_1$, \textcolor{black}{$9.2$}\% at $z_2$ and \textcolor{black}{$2.7$}\% at $z_3$, which shows a good agreement in high intensity areas.

\begin{figure}[ht!]
    \centering
    \includegraphics[width=0.48\textwidth]{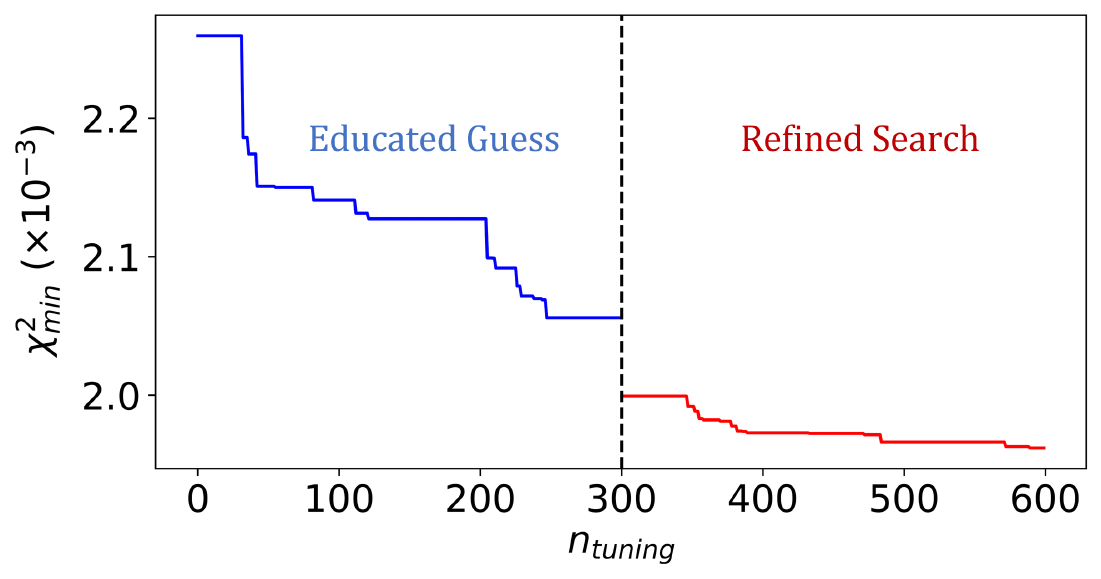}
    \caption{Evolution of the minimum error $\chi^2_{min}$ obtained during the center tuning of the GSA-MD applied to the LLC data-set as a function of the tuning iteration $n_{tuning}$, with $N_{iter}=50$ in Algorithm \ref{alg:field_reconstruction} for both the EG and RS phase. The blue curve is the evolution of $\chi^2_{min}$ in the EG phase with $N_{m}=N_{n}=10$ and the red curve is the evolution of $\chi^2_{min}$ in the RS phase with $N_{m}=N_{n}=30$.}
    \label{fig:chi_evolution}
\end{figure}

The evolution of the minimum error $\chi^2_{min}$ obtained during the center tuning is plotted for the EG and RS phases successively in Figure \ref{fig:chi_evolution}. The tuning of the HG centers leads to a reduction of $\chi^2_{min}$ from $2.26\times10^{-3}$ to $2.05\times10^{-3}$ during the EG phase, which corresponds to a $9$\% reduction. Using the optimized centers $(x_{0,k},y_{0,k})$ obtained with the EG as input of the RS yields $\chi_{min}^2=2.02\times10^{-3}$ at the start of RS the phase. This sudden reduction of $\chi^2_{min}$ between the end of the EG phase and the start of the RS phase is due to the higher number of HG modes used in the RS, which yields a more accurate field reconstruction and thus a lower $\chi^2_{min}$. The calculated HG coefficients at the end of the Refined Search can be used to quantify the degree of asymmetry of the data-set. For $N_m=N_n=10$, the partial sum $\sum_{m=0}^{N_m}\sum_{n=0}^{N_n}\left|C_{m,n}\right|^2$
reaches $97\%$ of the sum obtained using all HG coefficients.

During the RS, $\chi^2_{min}$ decreases from $\chi_{min}^2=2.02\times10^{-3}$ to $\chi_{min}^2=1.89\times10^{-3}$ , which corresponds to a \textcolor{black}{$6$}\% reduction. 
This shows that for this data-set, the EG alone is sufficient to find HG centers yielding a \textcolor{black}{minimized} error. 

It is important to use a high number of modes for a better reconstruction, as shown by the gap between the end of EG and start of RS.  To find the optimum centers with an RS phase, it may be necessary to adjust the parameters of the Bayesian Optimization itself to minimize the computational cost of the RS. 

For the LLC data-set, both the EG and RS phases to obtain the results presented in Figs. \ref{fig:HG_fit}, \ref{fig:itg_fit}, \ref{fig:chi_evolution} were performed on a laptop with CPU Intel i7-12700h, 64 GB RAM. The Bayesian Optimization phases were performed with 3 concurrent working threads.
In the EG phase, the required computing time was \textcolor{black}{19} minutes, and \textcolor{black}{42} minutes during the RS phase.

\begin{figure}[ht!]
    \centering
    \includegraphics[width=0.45\textwidth]{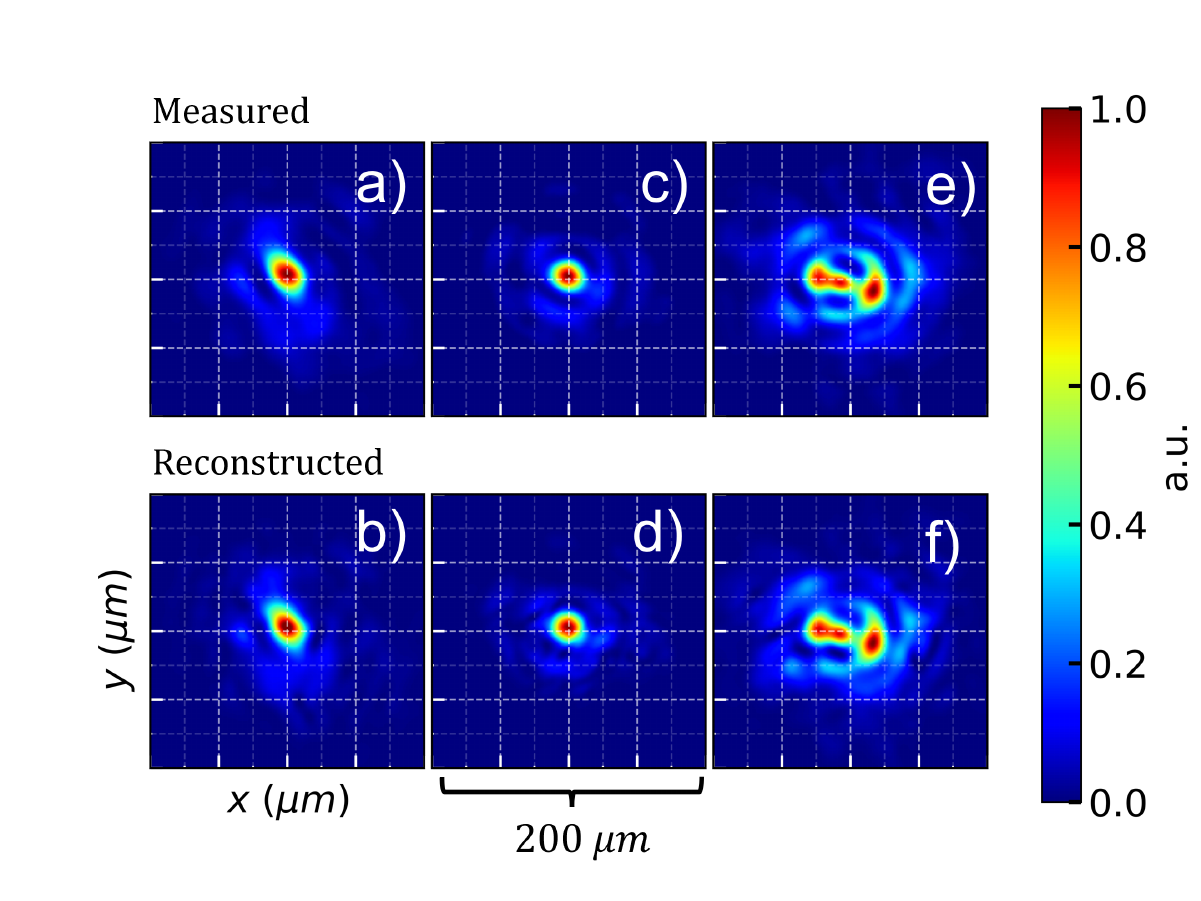}
    \caption{Measured fluence distribution of the Apollon data-set (upper row) and corresponding reconstructed distributions after the center tuning (lower row).  From left to right, the positions of the image planes along the propagation axis are :
    a, b) $z_1=-1800$ $\mu$m ($-1.8 \thinspace z_R$); c, d) $z_0=0$ $\mu$m; e, f) $z_1=1200$ $\mu$m ($1.2\thinspace z_R$). For each position $z_k$, the fluence has been normalized to the maximum of the corresponding measured fluence.}
    \label{fig:HG_fit_AP}
\end{figure}

\begin{figure}[ht!]
    \centering
    \includegraphics[width = 0.4\textwidth]{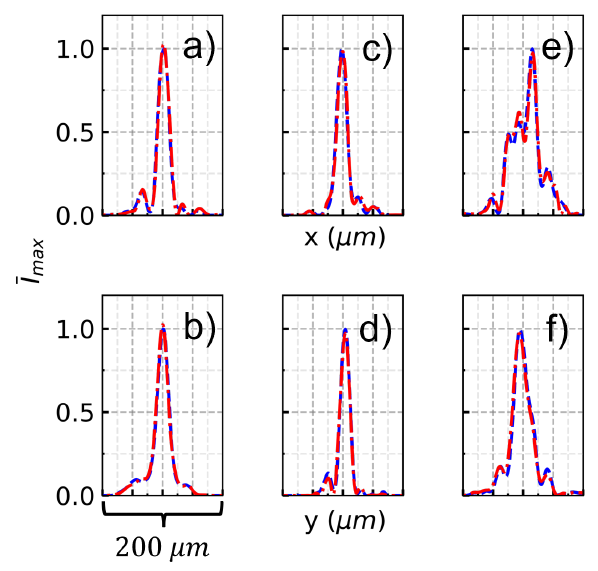}
    \caption{Fluence profiles along the $x$ (upper row) and $y$ (lower row) directions, respectively averaged over 3 pixels (2.55 $\mu$m) centered around the measured fluence maximum's position in $y$ and $x$. Each profile has been normalized to the measured fluence maximum at $z_k$. For a given position $z_k$, the blue dashed line is the measured fluence from the Apollon data-set and the red dashed line is the reconstructed fluence profile. From left to right, relative positions to the focal plane are :
    a, b) $z_1=-1800$ $\mu$m ($-1.8 \thinspace z_R$); c, d) $z_0=0$ $\mu$m; e, f) $z_1=1200$ $\mu$m ($1.2\thinspace z_R$).}
    \label{fig:itg_fit_AP}
\end{figure}

\begin{figure}[ht!]
    \centering
    \includegraphics[width=0.48\textwidth]{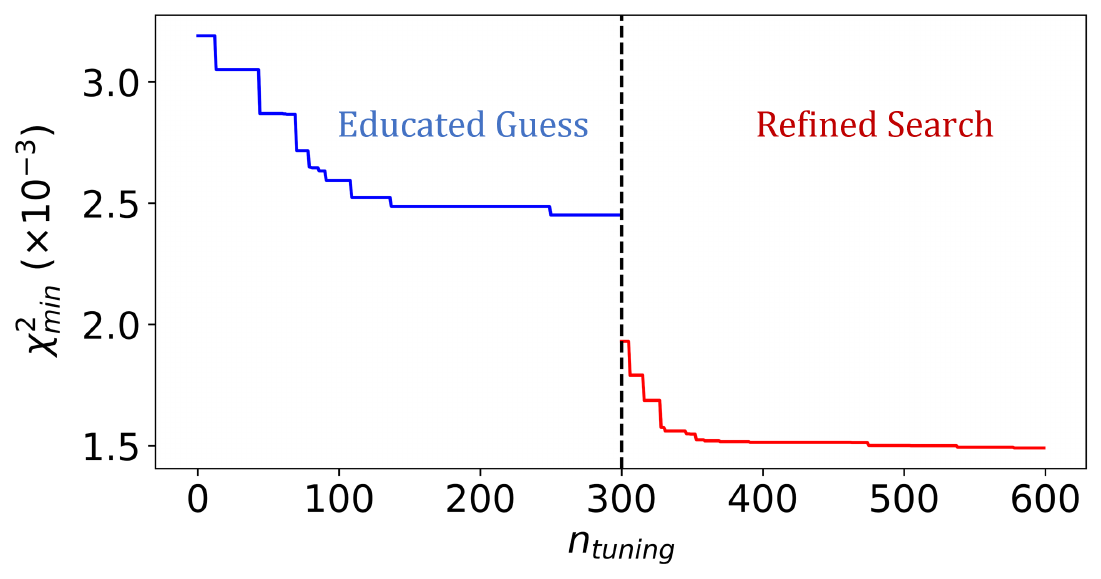}
    \caption{Evolution of the minimum error $\chi^2_{min}$ obtained during the center tuning of the GSA-MD applied to the Apollon data-set as a function of the tuning iteration $n_{tuning}$, with $N_{iter}=50$ in Algorithm \ref{alg:field_reconstruction} for both the EG and RS phase. The blue curve is the evolution of $\chi^2_{min}$ in the EG phase with $N_{m}=N_{n}=10$ and the red curve is the evolution of $\chi^2_{min}$ in the RS phase with $N_{m}=N_{n}=40$.}
    \label{fig:chi_evolution_AP}
\end{figure}

\subsection{Field reconstruction for the Apollon data-set}
For the Apollon data-set, the average shot energy is 4.8 J, for an average laser pulse duration of 25 fs, which represents a peak power $P_0=400$ TW. 
The central wavelength is $\lambda_0=0.8$ $\mu$m, and the waist of a Gaussian fit of the data measured in the focal plane is estimated at $w_{0,Gauss}=16$ $\mu$m, which sets the Rayleigh length of the Gaussian fit to $z_R\simeq1$ mm.

The Apollon data-set to reconstruct is a set of 3 individual transverse fluence distributions $F_{exp}(x,y,z_k)$ at $z_{0,1,2}=0$, $-1800$, $1200$ $\mu$m. Note that with this data-set the $z_0$ is the focal plane position, which is not the first position available on the propagation axis. Due to high shot to shot fluctuations, for a given position $z_k$, the fluence profile $F_{exp}(x,y,z_k)$ has been picked randomly among 4 images for $k=0$, and among 2 images for $k\neq 0$. The set of images over which the GSA-MD was performed is the same as in \cite{POP_IM}. The same process as the one used for the LLC data-set has been performed.

The same GSA-MD with Bayesian Optimization of the HG centers used for the LLC data-set was applied to the data of the Apollon Commissioning phase. %The effects of the pointing instability and asymmetry of these laser shots are discussed in \cite{POP_IM}. Pointing instabilities and laser pulse's transverse asymmetries play an important role in the LWFA physics investigated during the experiment; Taking into account the laser field reconstructed through the GSA-MD in LWFA simulations was shown to be of paramount importance to find a quantitative agreement between measured and simulated electron spectra. 
The size of the projecting grid was set at 301$\times$301 pixels, and number of modes in the RS phase to $N_n=N_m=40$. 

%Due to the limited size of the reconstruction grid and the dispersion of the centroids between the selected planes due to the pointing instability, each measured image has been centered around its individual fluence's centroid.

Compared to the LLC data-set, the relative pointing instability of the Apollon data-set is of the order of seven times larger (see Table \ref{tab:parameters}).
Thus, the search areas for the center tuning were chosen to be broader intervals compared to the search areas with the LLC data-set:
$S_{EG}=($ 100 $\mu$m$\times$ 100 $\mu$m$)$ centered around the centroid of the fluence distribution at $z=z_0$,  and  $S_{RS}=($ 20 $\mu$m$\times$ 20 $\mu$m$)$ centered around the calibrated centers found by the Educated Guess.
In both EG and RS phases, the number of iterations for the center tuning was set to $N_{tuning,EG} = N_{tuning,RS}=300$.

The results of the GSA-MD with HG centers optimization as well as the convergence of $\chi^2$ for the Apollon data-set are displayed in Figs. \ref{fig:HG_fit_AP}, \ref{fig:itg_fit_AP}, \ref{fig:chi_evolution_AP} respectively. For this application of the GSA-MD, again parallelized over 3 threads on the same laptop used with the LLC data-set, the EG phase took \textcolor{black}{18} minutes and the RS phase took \textcolor{black}{57} minutes. 

In Figure \ref{fig:HG_fit_AP}, the 2D comparison between the measured and reconstructed fluences shows a good agreement in the energy distribution of measurements and reconstructions.

In Figure \ref{fig:itg_fit_AP}, the comparison between measured 1D profiles and reconstructed profiles at the measured fluence's maximum shows a good agreement in the amplitude. The maximum relative differences on the measured fluence's amplitude in x and y are \textcolor{black}{$2.7$}\% at $z_0$, \textcolor{black}{$2.9$}\% at $z_1$ and \textcolor{black}{$0.8$}\% at $z_2$. 

In Figure \ref{fig:chi_evolution_AP}, the evolution of the minimum error $\chi^2_{min}$ over the center tuning process is reported. 
The relative $\chi^2_{min}$ gap when going from the EG to RS phase at $n_{tuning}=300$ is larger than for the results with the LLC data-set (see Fig. \ref{fig:chi_evolution}), due to the greater difference in the number of HG modes used in the EG and RS phase. \textcolor{black}{For the Apollon data-set, setting $N_m=N_n=10$, the partial sum $\sum_{m=0}^{N_m}\sum_{n=0}^{N_n}\left|C_{m,n}\right|^2$
reaches only $90\%$ of the sum obtained using all HG coefficients, while for the LLC data-set this number reaches $97\%$.}
This highlights the importance of using a high number of HG modes used for the GSA-MD calculation, especially in the RS. In this later phase, $\chi^2_{min}$ is decreased by \textcolor{black}{$18\%$}, which is \textcolor{black}{on par with the decrease of} the EG (\textcolor{black}{$23\%$}).
\textcolor{black}{In comparison to the LLC data-set, the Refined Search phase of the Apollon data-set GSA-MD has a quicker convergence of the reconstruction error. The difference stems from a higher sum share when fixing $N_{m}$, $N_{n}=10$ for the LLC data-set.}

\textcolor{black}{\subsection{Comparison with a version of the Gerchberg-Saxton algorithm without modes decomposition}}

\textcolor{black}{
In this section the performances of the GSA-MD are compared to those of a version of the GSA that uses the Fresnel Transform for the propagation of the electric field \cite{zalevsky1996gerchberg}. The flowchart of this implementation is the same as in the 3D Gerchberg-Saxton variant of \cite{Zhou2019}, except for the amplitude constraint which here is Step 5) of Algorithm \ref{alg:field_reconstruction}. To compare the results of the GSA-MD with this GSA version (for brevity referred to as "GSA" in the following), the Apollon data-set was used. 
The GSA has been performed with $z_0$ defined as the reference plane, and $z_{1,2}$ as the image planes.
The GSA-MD has been performed with $N_{m}=N_{n}=40$ and without origin tuning, and with $N_{m}=N_{n}=40$ and origin tuning. The same maximum number of iterations, i.e. $N_{iter}=50$ was set for the GSA and for the Algorithm \ref{alg:field_reconstruction} for the GSA-MD.}

\begin{figure}[ht!]
    \centering
    \includegraphics[width = 0.45\textwidth]{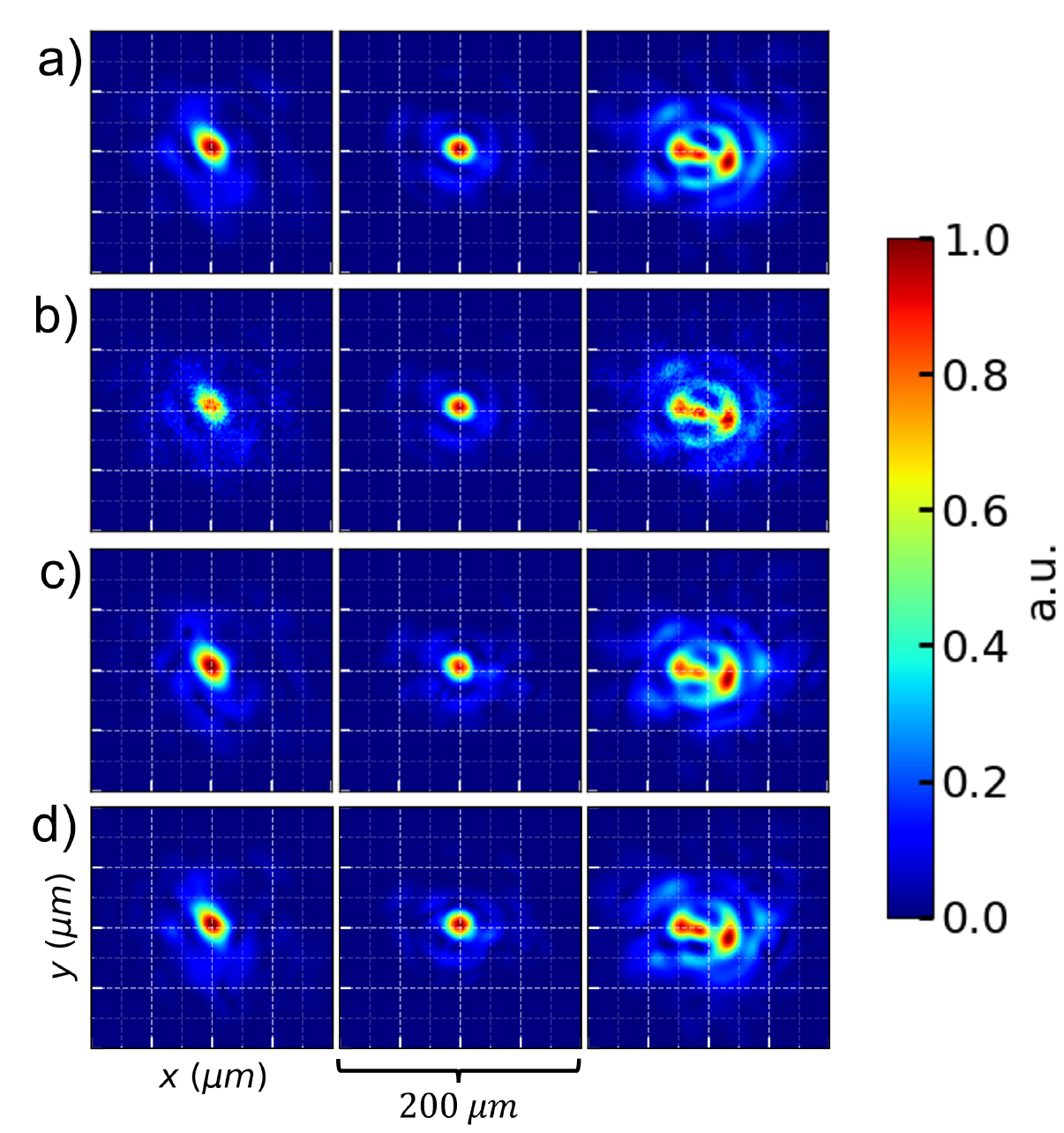}
    \caption{\textcolor{black}{Reconstructed distributions for the Apollon data-set.
    a) measured fluence distributions; b) fluence distributions reconstructed with the GSA; c) fluence distributions reconstructed with the GSA-MD and $N_m=N_n=40$, without origin tuning; d) fluence distributions reconstructed with the GSA-MD, $N_m=N_n=40$ and origin tuning.\\
    From left to right, the positions of the image planes along the propagation axis are :
     $z_1=-1800$ $\mu$m ($-1.8 \thinspace z_R$), left column; $z_0=0$ $\mu$m, middle column; $z_1=1200$ $\mu$m ($1.2\thinspace z_R$), right column.}}
    \label{fig:HG_comparison}
\end{figure}

\textcolor{black}{The results for the GSA and the 2 runs of GSA-MD (without and with origin tuning) are displayed in Fig. \ref{fig:HG_comparison}. Although the reconstructions displayed in Figs. \ref{fig:HG_comparison}.(b)-(d) are qualitatively similar, the reconstructed fluence distributions obtained with the GSA in $z_1$ and $z_2$ of Fig. \ref{fig:HG_comparison}.(b) are noisier than the ones from Figs. \ref{fig:HG_comparison}.(c) and (d) obtained with the GSA-MD.
To quantify this noise across the planes $z_k$, the error $\chi^2_k$ was measured for each plane. It is defined as :}

\textcolor{black}{
\begin{equation}
\label{eq:error_k}
\chi^2_k =\dfrac{\sqrt{\sum_{i_x,\thinspace i_y}^{N_{pix_x},N_{pix_y}} \left( F_{exp}(x,y,z_k)-F_{fit}(x,y,z_k)\right)^2}}{{\sum_{i_x,\thinspace i_y}^{N_{pix_x},N_{pix_y}} F_{exp}(x,y,z_k)}}.
\end{equation}
}

\textcolor{black}{
By definition $\chi^2$ defined in Eq. \ref{eq:error_chi} is the average of the errors $\chi^2_k$ of all planes, i.e. $\chi^2 = \dfrac{1}{N_{images}}\sum_{k=0}^{N_{images}-1} \chi^2_k$.}

\textcolor{black}{The performances of the GSA and of the GSA-MD without and with origin tuning are reported in Table \ref{tab:performances}. Note that some of the data reported in the third column of Table \ref{tab:performances} appear in the third column of Table \ref{tab:parameters}.}

\begin{table}
\small
\centering
\caption{Performances on the Apollon data-set of the GSA and of the GSA-MD without and with origin tuning.
The value $\chi^2_k$ is the value of the reconstruction error in the plane $z_k$.\\
Note that the reported total times for the GSA-MD were obtained using the stopping criterion on the error gradient in Algorithm \ref{alg:field_reconstruction}.
For the GSA-MD without origin tuning, $iter_{break}=45$. With origin tuning, this value varies at each origin tuning iteration.}
\resizebox{\linewidth}{!}{\begin{tabular}{cccc}
\hline
Parameter & GSA & GSA-MD  & GSA-MD \\
 & & ($N_{m,n}=40$, & ($N_{m,n}=40$,\\
 & & without origin tuning) & with origin tuning)\\
\hline
$N_{iter}$ & 50 & 50 & 50 \\
Total time & 3.8 s & 13.6 s & 1h15 min\\
$\chi^2$ ($\times 10^{-3}$) & $2.50$ & $2.28$ & $1.61$\\
$\chi^2_0$ ($\times 10^{-3}$) & $1.94$ & $2.76$ & $1.98$\\
$\chi^2_1$ ($\times 10^{-3}$)& $3.89$ & $2.48$ & $1.52$\\
$\chi^2_2$($\times 10^{-3}$) & $1.67$ & $1.60$ & $1.35$\\
\hline
\end{tabular}}\label{tab:performances}
\end{table}

\textcolor{black}{
The GSA-MD with $N_m=N_n=40$ and no origin tuning yields a $\chi^2$ error $9\%$ lower than the GSA variant. With the origin tuning, the $\chi^2$ error of the GSA-MD becomes $35\%$ lower. Furthermore, the reconstructed profiles in $z_1$ and $z_2$ of Fig. 11.(b) are noisier than their GSA-MD counterparts. This difference results into higher values of $\chi^2_{1}$ and $\chi^2_{2}$.
The difference between the maximum $\chi^2_k$ and the minimum $\chi^2_k$ across the planes is equal to $89\%$, $51\%$, $39\%$ of the average error $\chi^2$ for the GSA, GSA-MD without and with origin tuning respectively.}

\textcolor{black}{To summarize, the GSA-MD without origin tuning and $N_m=N_n=40$ has an execution time of the order of ten seconds, while the GSA has an execution time of 3.8 s. With $N_m=N_n=10$, the GSA-MD without origin tuning performs in a shorter execution time of 2.7 s and $\chi^2=3.2\times10^{-3}$ (this case is not included in Table \ref{tab:performances} and Fig. \ref{fig:HG_comparison}). Additionally, the considered GSA-MD results with $N_m=N_n=40$ yield a lower reconstruction error, a more uniform distribution of the reconstruction errors $\chi^2_k$ across the planes, and smoother distributions in $z_{1,2}$. Using the origin tuning in GSA-MD makes the distribution of the reconstruction errors $\chi^2_k$ even more uniform across the planes. }
\\

\section{Conclusions}
\textcolor{black}{A fast, flexible Gerchberg-Saxton algorithm with Hermite-Gauss mode decomposition to reconstruct the laser field was presented.
In this algorithm, as in a 3D Gerchberg-Saxton Algorithm, the fluence data from multiple planes is used to iteratively build a description of the laser pulse (amplitude and phase). This knowledge can be used to study, and possibly correct, the imperfections of high intensity laser pulses and their effect in laser-plasma interaction.}

Compared to a Gerchberg-Saxton algorithm using propagators of Fourier transforms, the use of modes in the proposed algorithm introduces some flexibility.
Since the measured fluences come from different shots, often with wavefront and pointing instabilities, tuning the centers of the modes allows to reduce the error associated to the field reconstruction. 
Changing the number of modes allows to reach the desired compromise between reconstruction error and required computation time for the reconstruction.

These features of the algorithm have been demonstrated showing the reconstruction of the laser field of two very different high intensity lasers, the Lund Laser Centre (LLC) system and the Apollon facility in the commissioning phase. The results of the presented algorithm with the two data-sets display a good agreement between  measured and  reconstructed fluences. 
%\textcolor{black}{It was also noted during the data-sets analysis that if an averaging operation is done on images at the same plane, the reconstruction algorithm can yield a reduced $\chi^2$ value.}
\textcolor{black}{The reconstruction of the electric field needed approximately 1 hour and 1 hour 15 minutes on a laptop for the LLC, and Apollon data-sets, respectively. It has been shown that with the Apollon data-set and using 40 HG modes in both directions, the GSA-MD can yield a field reconstruction less noisy than a Gerchberg-Saxton algorithm without modes decomposition. In this comparison, a smaller reconstruction error and a more uniform distribution of this error across the planes were obtained, both with and without origin tuning. Without origin tuning, the GSA-MD with 40 HG modes can have an execution time of the order of ten seconds, and of a few seconds with a lower number of modes.}

The presented algorithm can thus become a valuable tool for the study, and possibly the correction in the long term,  of the transverse imperfections of high intensity laser systems with femtosecond pulses.

\section{Acknowledgements}
Experimental data were collected during an experimental campaign at the Lund Laser Centre, which received funding from the European Union’s Horizon 2020 Research and Innovation Programme under Grant Agreement No. 730871 and during an experimental campaign at Apollon Research Infrastructure, partially funded by Equipex Cilex (Centre interdisciplinaire lumière extrême) grant N° ANR-10-EQPX-25-01, and by region Ile-de-France.

% Bibliography
\bibliography{Bibliography}

\begin{thebibliography}{43}%
\makeatletter
\providecommand \@ifxundefined [1]{%
 \@ifx{#1\undefined}
}%
\providecommand \@ifnum [1]{%
 \ifnum #1\expandafter \@firstoftwo
 \else \expandafter \@secondoftwo
 \fi
}%
\providecommand \@ifx [1]{%
 \ifx #1\expandafter \@firstoftwo
 \else \expandafter \@secondoftwo
 \fi
}%
\providecommand \natexlab [1]{#1}%
\providecommand \enquote  [1]{``#1''}%
\providecommand \bibnamefont  [1]{#1}%
\providecommand \bibfnamefont [1]{#1}%
\providecommand \citenamefont [1]{#1}%
\providecommand \href@noop [0]{\@secondoftwo}%
\providecommand \href [0]{\begingroup \@sanitize@url \@href}%
\providecommand \@href[1]{\@@startlink{#1}\@@href}%
\providecommand \@@href[1]{\endgroup#1\@@endlink}%
\providecommand \@sanitize@url [0]{\catcode `\\12\catcode `\$12\catcode
  `\&12\catcode `\#12\catcode `\^12\catcode `\_12\catcode `\%12\relax}%
\providecommand \@@startlink[1]{}%
\providecommand \@@endlink[0]{}%
\providecommand \url  [0]{\begingroup\@sanitize@url \@url }%
\providecommand \@url [1]{\endgroup\@href {#1}{\urlprefix }}%
\providecommand \urlprefix  [0]{URL }%
\providecommand \Eprint [0]{\href }%
\providecommand \doibase [0]{https://doi.org/}%
\providecommand \selectlanguage [0]{\@gobble}%
\providecommand \bibinfo  [0]{\@secondoftwo}%
\providecommand \bibfield  [0]{\@secondoftwo}%
\providecommand \translation [1]{[#1]}%
\providecommand \BibitemOpen [0]{}%
\providecommand \bibitemStop [0]{}%
\providecommand \bibitemNoStop [0]{.\EOS\space}%
\providecommand \EOS [0]{\spacefactor3000\relax}%
\providecommand \BibitemShut  [1]{\csname bibitem#1\endcsname}%
\let\auto@bib@innerbib\@empty
%</preamble>
\bibitem [{\citenamefont {Strickland}\ and\ \citenamefont
  {Mourou}(1985)}]{Strickland1985}%
  \BibitemOpen
  \bibfield  {author} {\bibinfo {author} {\bibfnamefont {D.}~\bibnamefont
  {Strickland}}\ and\ \bibinfo {author} {\bibfnamefont {G.}~\bibnamefont
  {Mourou}},\ }\href
  {https://doi.org/https://doi.org/10.1016/0030-4018(85)90120-8} {\bibfield
  {journal} {\bibinfo  {journal} {Optics Communications}\ }\textbf {\bibinfo
  {volume} {56}},\ \bibinfo {pages} {219} (\bibinfo {year} {1985})}\BibitemShut
  {NoStop}%
\bibitem [{\citenamefont {Ranc}\ \emph {et~al.}(2000)\citenamefont {Ranc},
  \citenamefont {Ch{\'e}riaux}, \citenamefont {Ferr{\'e}}, \citenamefont
  {Rousseau},\ and\ \citenamefont {Chambaret}}]{Ranc2000}%
  \BibitemOpen
  \bibfield  {author} {\bibinfo {author} {\bibfnamefont {S.}~\bibnamefont
  {Ranc}}, \bibinfo {author} {\bibfnamefont {G.}~\bibnamefont {Ch{\'e}riaux}},
  \bibinfo {author} {\bibfnamefont {S.}~\bibnamefont {Ferr{\'e}}}, \bibinfo
  {author} {\bibfnamefont {J.~P.}\ \bibnamefont {Rousseau}},\ and\ \bibinfo
  {author} {\bibfnamefont {J.~P.}\ \bibnamefont {Chambaret}},\ }\href
  {https://doi.org/10.1007/s003400000318} {\bibfield  {journal} {\bibinfo
  {journal} {Applied Physics B}\ }\textbf {\bibinfo {volume} {70}},\ \bibinfo
  {pages} {S181} (\bibinfo {year} {2000})}\BibitemShut {NoStop}%
\bibitem [{\citenamefont {Yoon}\ \emph {et~al.}(2021)\citenamefont {Yoon},
  \citenamefont {Kim}, \citenamefont {Choi}, \citenamefont {Sung},
  \citenamefont {Lee}, \citenamefont {Lee},\ and\ \citenamefont
  {Nam}}]{Yoon2021}%
  \BibitemOpen
  \bibfield  {author} {\bibinfo {author} {\bibfnamefont {J.~W.}\ \bibnamefont
  {Yoon}}, \bibinfo {author} {\bibfnamefont {Y.~G.}\ \bibnamefont {Kim}},
  \bibinfo {author} {\bibfnamefont {I.~W.}\ \bibnamefont {Choi}}, \bibinfo
  {author} {\bibfnamefont {J.~H.}\ \bibnamefont {Sung}}, \bibinfo {author}
  {\bibfnamefont {H.~W.}\ \bibnamefont {Lee}}, \bibinfo {author} {\bibfnamefont
  {S.~K.}\ \bibnamefont {Lee}},\ and\ \bibinfo {author} {\bibfnamefont {C.~H.}\
  \bibnamefont {Nam}},\ }\href {https://doi.org/10.1364/OPTICA.420520}
  {\bibfield  {journal} {\bibinfo  {journal} {Optica}\ }\textbf {\bibinfo
  {volume} {8}},\ \bibinfo {pages} {630} (\bibinfo {year} {2021})}\BibitemShut
  {NoStop}%
\bibitem [{\citenamefont {Dickson}\ \emph {et~al.}(2022)\citenamefont
  {Dickson}, \citenamefont {Underwood}, \citenamefont {Filippi}, \citenamefont
  {Shalloo}, \citenamefont {Svensson}, \citenamefont {Gu\'enot}, \citenamefont
  {Svendsen}, \citenamefont {Moulanier}, \citenamefont {Dufr\'enoy},
  \citenamefont {Murphy}, \citenamefont {Lopes}, \citenamefont {Rajeev},
  \citenamefont {Najmudin}, \citenamefont {Cantono}, \citenamefont {Persson},
  \citenamefont {Lundh}, \citenamefont {Maynard}, \citenamefont {Streeter},\
  and\ \citenamefont {Cros}}]{Dickson2022}%
  \BibitemOpen
  \bibfield  {author} {\bibinfo {author} {\bibfnamefont {L.~T.}\ \bibnamefont
  {Dickson}}, \bibinfo {author} {\bibfnamefont {C.~I.~D.}\ \bibnamefont
  {Underwood}}, \bibinfo {author} {\bibfnamefont {F.}~\bibnamefont {Filippi}},
  \bibinfo {author} {\bibfnamefont {R.~J.}\ \bibnamefont {Shalloo}}, \bibinfo
  {author} {\bibfnamefont {J.~B.}\ \bibnamefont {Svensson}}, \bibinfo {author}
  {\bibfnamefont {D.}~\bibnamefont {Gu\'enot}}, \bibinfo {author}
  {\bibfnamefont {K.}~\bibnamefont {Svendsen}}, \bibinfo {author}
  {\bibfnamefont {I.}~\bibnamefont {Moulanier}}, \bibinfo {author}
  {\bibfnamefont {S.~D.}\ \bibnamefont {Dufr\'enoy}}, \bibinfo {author}
  {\bibfnamefont {C.~D.}\ \bibnamefont {Murphy}}, \bibinfo {author}
  {\bibfnamefont {N.~C.}\ \bibnamefont {Lopes}}, \bibinfo {author}
  {\bibfnamefont {P.~P.}\ \bibnamefont {Rajeev}}, \bibinfo {author}
  {\bibfnamefont {Z.}~\bibnamefont {Najmudin}}, \bibinfo {author}
  {\bibfnamefont {G.}~\bibnamefont {Cantono}}, \bibinfo {author} {\bibfnamefont
  {A.}~\bibnamefont {Persson}}, \bibinfo {author} {\bibfnamefont
  {O.}~\bibnamefont {Lundh}}, \bibinfo {author} {\bibfnamefont
  {G.}~\bibnamefont {Maynard}}, \bibinfo {author} {\bibfnamefont {M.~J.~V.}\
  \bibnamefont {Streeter}},\ and\ \bibinfo {author} {\bibfnamefont
  {B.}~\bibnamefont {Cros}},\ }\href
  {https://doi.org/10.1103/PhysRevAccelBeams.25.101301} {\bibfield  {journal}
  {\bibinfo  {journal} {Phys. Rev. Accel. Beams}\ }\textbf {\bibinfo {volume}
  {25}},\ \bibinfo {pages} {101301} (\bibinfo {year} {2022})}\BibitemShut
  {NoStop}%
\bibitem [{\citenamefont {Moulanier}\ \emph {et~al.}(2023)\citenamefont
  {Moulanier}, \citenamefont {Dickson}, \citenamefont {Ballage}, \citenamefont
  {Vasilovici}, \citenamefont {Gremaud}, \citenamefont {Dobosz~Dufr{\'e}noy},
  \citenamefont {Delerue}, \citenamefont {Bernardi}, \citenamefont {Mahjoub},
  \citenamefont {Cauchois} \emph {et~al.}}]{POP_IM}%
  \BibitemOpen
  \bibfield  {author} {\bibinfo {author} {\bibfnamefont {I.}~\bibnamefont
  {Moulanier}}, \bibinfo {author} {\bibfnamefont {L.}~\bibnamefont {Dickson}},
  \bibinfo {author} {\bibfnamefont {C.}~\bibnamefont {Ballage}}, \bibinfo
  {author} {\bibfnamefont {O.}~\bibnamefont {Vasilovici}}, \bibinfo {author}
  {\bibfnamefont {A.}~\bibnamefont {Gremaud}}, \bibinfo {author} {\bibfnamefont
  {S.}~\bibnamefont {Dobosz~Dufr{\'e}noy}}, \bibinfo {author} {\bibfnamefont
  {N.}~\bibnamefont {Delerue}}, \bibinfo {author} {\bibfnamefont
  {L.}~\bibnamefont {Bernardi}}, \bibinfo {author} {\bibfnamefont
  {A.}~\bibnamefont {Mahjoub}}, \bibinfo {author} {\bibfnamefont
  {A.}~\bibnamefont {Cauchois}}, \emph {et~al.},\ }\href@noop {} {\bibfield
  {journal} {\bibinfo  {journal} {Physics of Plasmas}\ }\textbf {\bibinfo
  {volume} {30}} (\bibinfo {year} {2023})}\BibitemShut {NoStop}%
\bibitem [{\citenamefont {Santarsiero}\ \emph {et~al.}(1997)\citenamefont
  {Santarsiero}, \citenamefont {Aiello}, \citenamefont {Borghi},\ and\
  \citenamefont {Vicalvi}}]{Santarsiero1997}%
  \BibitemOpen
  \bibfield  {author} {\bibinfo {author} {\bibfnamefont {M.}~\bibnamefont
  {Santarsiero}}, \bibinfo {author} {\bibfnamefont {D.}~\bibnamefont {Aiello}},
  \bibinfo {author} {\bibfnamefont {R.}~\bibnamefont {Borghi}},\ and\ \bibinfo
  {author} {\bibfnamefont {S.}~\bibnamefont {Vicalvi}},\ }\href
  {https://doi.org/10.1080/09500349708232927} {\bibfield  {journal} {\bibinfo
  {journal} {Journal of Modern Optics}\ }\textbf {\bibinfo {volume} {44}},\
  \bibinfo {pages} {633} (\bibinfo {year} {1997})},\ \Eprint
  {https://arxiv.org/abs/https://doi.org/10.1080/09500349708232927}
  {https://doi.org/10.1080/09500349708232927} \BibitemShut {NoStop}%
\bibitem [{\citenamefont {Akturk}\ \emph {et~al.}(2010)\citenamefont {Akturk},
  \citenamefont {Gu}, \citenamefont {Bowlan},\ and\ \citenamefont
  {Trebino}}]{Akturk2010}%
  \BibitemOpen
  \bibfield  {author} {\bibinfo {author} {\bibfnamefont {S.}~\bibnamefont
  {Akturk}}, \bibinfo {author} {\bibfnamefont {X.}~\bibnamefont {Gu}}, \bibinfo
  {author} {\bibfnamefont {P.}~\bibnamefont {Bowlan}},\ and\ \bibinfo {author}
  {\bibfnamefont {R.}~\bibnamefont {Trebino}},\ }\href
  {https://doi.org/10.1088/2040-8978/12/9/093001} {\bibfield  {journal}
  {\bibinfo  {journal} {Journal of Optics}\ }\textbf {\bibinfo {volume} {12}},\
  \bibinfo {pages} {093001} (\bibinfo {year} {2010})}\BibitemShut {NoStop}%
\bibitem [{\citenamefont {Jeandet}\ \emph {et~al.}(2022)\citenamefont
  {Jeandet}, \citenamefont {Jolly}, \citenamefont {Borot}, \citenamefont
  {Bussi\`{e}re}, \citenamefont {Dumont}, \citenamefont {Gautier},
  \citenamefont {Gobert}, \citenamefont {Goddet}, \citenamefont {Gonsalves},
  \citenamefont {Irman}, \citenamefont {Leemans}, \citenamefont
  {Lopez-Martens}, \citenamefont {Mennerat}, \citenamefont {Nakamura},
  \citenamefont {Ouill\'{e}}, \citenamefont {Pariente}, \citenamefont
  {Pittman}, \citenamefont {P\"{u}schel}, \citenamefont {Sanson}, \citenamefont
  {Sylla}, \citenamefont {Thaury}, \citenamefont {Zeil},\ and\ \citenamefont
  {Qu\'{e}r\'{e}}}]{Jeandet2022}%
  \BibitemOpen
  \bibfield  {author} {\bibinfo {author} {\bibfnamefont {A.}~\bibnamefont
  {Jeandet}}, \bibinfo {author} {\bibfnamefont {S.~W.}\ \bibnamefont {Jolly}},
  \bibinfo {author} {\bibfnamefont {A.}~\bibnamefont {Borot}}, \bibinfo
  {author} {\bibfnamefont {B.}~\bibnamefont {Bussi\`{e}re}}, \bibinfo {author}
  {\bibfnamefont {P.}~\bibnamefont {Dumont}}, \bibinfo {author} {\bibfnamefont
  {J.}~\bibnamefont {Gautier}}, \bibinfo {author} {\bibfnamefont
  {O.}~\bibnamefont {Gobert}}, \bibinfo {author} {\bibfnamefont {J.-P.}\
  \bibnamefont {Goddet}}, \bibinfo {author} {\bibfnamefont {A.}~\bibnamefont
  {Gonsalves}}, \bibinfo {author} {\bibfnamefont {A.}~\bibnamefont {Irman}},
  \bibinfo {author} {\bibfnamefont {W.~P.}\ \bibnamefont {Leemans}}, \bibinfo
  {author} {\bibfnamefont {R.}~\bibnamefont {Lopez-Martens}}, \bibinfo {author}
  {\bibfnamefont {G.}~\bibnamefont {Mennerat}}, \bibinfo {author}
  {\bibfnamefont {K.}~\bibnamefont {Nakamura}}, \bibinfo {author}
  {\bibfnamefont {M.}~\bibnamefont {Ouill\'{e}}}, \bibinfo {author}
  {\bibfnamefont {G.}~\bibnamefont {Pariente}}, \bibinfo {author}
  {\bibfnamefont {M.}~\bibnamefont {Pittman}}, \bibinfo {author} {\bibfnamefont
  {T.}~\bibnamefont {P\"{u}schel}}, \bibinfo {author} {\bibfnamefont
  {F.}~\bibnamefont {Sanson}}, \bibinfo {author} {\bibfnamefont
  {F.}~\bibnamefont {Sylla}}, \bibinfo {author} {\bibfnamefont
  {C.}~\bibnamefont {Thaury}}, \bibinfo {author} {\bibfnamefont
  {K.}~\bibnamefont {Zeil}},\ and\ \bibinfo {author} {\bibfnamefont
  {F.}~\bibnamefont {Qu\'{e}r\'{e}}},\ }\href
  {https://doi.org/10.1364/OE.444564} {\bibfield  {journal} {\bibinfo
  {journal} {Opt. Express}\ }\textbf {\bibinfo {volume} {30}},\ \bibinfo
  {pages} {3262} (\bibinfo {year} {2022})}\BibitemShut {NoStop}%
\bibitem [{\citenamefont {Bourassin-Bouchet}\ \emph {et~al.}(2011)\citenamefont
  {Bourassin-Bouchet}, \citenamefont {Stephens}, \citenamefont {de~Rossi},
  \citenamefont {Delmotte},\ and\ \citenamefont
  {Chavel}}]{Bourassin-Bouchet2011}%
  \BibitemOpen
  \bibfield  {author} {\bibinfo {author} {\bibfnamefont {C.}~\bibnamefont
  {Bourassin-Bouchet}}, \bibinfo {author} {\bibfnamefont {M.}~\bibnamefont
  {Stephens}}, \bibinfo {author} {\bibfnamefont {S.}~\bibnamefont {de~Rossi}},
  \bibinfo {author} {\bibfnamefont {F.}~\bibnamefont {Delmotte}},\ and\
  \bibinfo {author} {\bibfnamefont {P.}~\bibnamefont {Chavel}},\ }\href
  {https://doi.org/10.1364/OE.19.017357} {\bibfield  {journal} {\bibinfo
  {journal} {Opt. Express}\ }\textbf {\bibinfo {volume} {19}},\ \bibinfo
  {pages} {17357} (\bibinfo {year} {2011})}\BibitemShut {NoStop}%
\bibitem [{\citenamefont {Li}\ \emph {et~al.}(2017)\citenamefont {Li},
  \citenamefont {Tsubakimoto}, \citenamefont {Yoshida}, \citenamefont
  {Nakata},\ and\ \citenamefont {Miyanaga}}]{Li2017}%
  \BibitemOpen
  \bibfield  {author} {\bibinfo {author} {\bibfnamefont {Z.}~\bibnamefont
  {Li}}, \bibinfo {author} {\bibfnamefont {K.}~\bibnamefont {Tsubakimoto}},
  \bibinfo {author} {\bibfnamefont {H.}~\bibnamefont {Yoshida}}, \bibinfo
  {author} {\bibfnamefont {Y.}~\bibnamefont {Nakata}},\ and\ \bibinfo {author}
  {\bibfnamefont {N.}~\bibnamefont {Miyanaga}},\ }\href
  {https://doi.org/10.7567/APEX.10.102702} {\bibfield  {journal} {\bibinfo
  {journal} {Applied Physics Express}\ }\textbf {\bibinfo {volume} {10}},\
  \bibinfo {pages} {102702} (\bibinfo {year} {2017})}\BibitemShut {NoStop}%
\bibitem [{\citenamefont {Li}\ and\ \citenamefont {Miyanaga}(2018)}]{Li2018}%
  \BibitemOpen
  \bibfield  {author} {\bibinfo {author} {\bibfnamefont {Z.}~\bibnamefont
  {Li}}\ and\ \bibinfo {author} {\bibfnamefont {N.}~\bibnamefont {Miyanaga}},\
  }\href {https://doi.org/10.1364/OE.26.008453} {\bibfield  {journal} {\bibinfo
   {journal} {Opt. Express}\ }\textbf {\bibinfo {volume} {26}},\ \bibinfo
  {pages} {8453} (\bibinfo {year} {2018})}\BibitemShut {NoStop}%
\bibitem [{\citenamefont {Fourmaux}\ \emph {et~al.}(2008)\citenamefont
  {Fourmaux}, \citenamefont {Payeur}, \citenamefont {Alexandrov}, \citenamefont
  {Serbanescu}, \citenamefont {Martin}, \citenamefont {Ozaki}, \citenamefont
  {Kudryashov},\ and\ \citenamefont {Kieffer}}]{Fourmaux2008}%
  \BibitemOpen
  \bibfield  {author} {\bibinfo {author} {\bibfnamefont {S.}~\bibnamefont
  {Fourmaux}}, \bibinfo {author} {\bibfnamefont {S.}~\bibnamefont {Payeur}},
  \bibinfo {author} {\bibfnamefont {A.}~\bibnamefont {Alexandrov}}, \bibinfo
  {author} {\bibfnamefont {C.}~\bibnamefont {Serbanescu}}, \bibinfo {author}
  {\bibfnamefont {F.}~\bibnamefont {Martin}}, \bibinfo {author} {\bibfnamefont
  {T.}~\bibnamefont {Ozaki}}, \bibinfo {author} {\bibfnamefont
  {A.}~\bibnamefont {Kudryashov}},\ and\ \bibinfo {author} {\bibfnamefont
  {J.~C.}\ \bibnamefont {Kieffer}},\ }\href
  {https://doi.org/10.1364/OE.16.011987} {\bibfield  {journal} {\bibinfo
  {journal} {Opt. Express}\ }\textbf {\bibinfo {volume} {16}},\ \bibinfo
  {pages} {11987} (\bibinfo {year} {2008})}\BibitemShut {NoStop}%
\bibitem [{\citenamefont {Zemzemi}\ \emph {et~al.}(2020)\citenamefont
  {Zemzemi}, \citenamefont {Massimo},\ and\ \citenamefont
  {Beck}}]{Zemzemi_2020}%
  \BibitemOpen
  \bibfield  {author} {\bibinfo {author} {\bibfnamefont {I.}~\bibnamefont
  {Zemzemi}}, \bibinfo {author} {\bibfnamefont {F.}~\bibnamefont {Massimo}},\
  and\ \bibinfo {author} {\bibfnamefont {A.}~\bibnamefont {Beck}},\ }\href
  {https://doi.org/10.1088/1742-6596/1596/1/012054} {\bibfield  {journal}
  {\bibinfo  {journal} {Journal of Physics: Conference Series}\ }\textbf
  {\bibinfo {volume} {1596}},\ \bibinfo {pages} {012054} (\bibinfo {year}
  {2020})}\BibitemShut {NoStop}%
\bibitem [{\citenamefont {Wodzinski}\ \emph {et~al.}(2020)\citenamefont
  {Wodzinski}, \citenamefont {K\"{u}nzel}, \citenamefont {Koliyadu},
  \citenamefont {Hussain}, \citenamefont {Keitel}, \citenamefont {Williams},
  \citenamefont {Zeitoun}, \citenamefont {Pl\"{o}njes},\ and\ \citenamefont
  {Fajardo}}]{Wodzinski2020}%
  \BibitemOpen
  \bibfield  {author} {\bibinfo {author} {\bibfnamefont {T.}~\bibnamefont
  {Wodzinski}}, \bibinfo {author} {\bibfnamefont {S.}~\bibnamefont
  {K\"{u}nzel}}, \bibinfo {author} {\bibfnamefont {J.~C.~P.}\ \bibnamefont
  {Koliyadu}}, \bibinfo {author} {\bibfnamefont {M.}~\bibnamefont {Hussain}},
  \bibinfo {author} {\bibfnamefont {B.}~\bibnamefont {Keitel}}, \bibinfo
  {author} {\bibfnamefont {G.~O.}\ \bibnamefont {Williams}}, \bibinfo {author}
  {\bibfnamefont {P.}~\bibnamefont {Zeitoun}}, \bibinfo {author} {\bibfnamefont
  {E.}~\bibnamefont {Pl\"{o}njes}},\ and\ \bibinfo {author} {\bibfnamefont
  {M.}~\bibnamefont {Fajardo}},\ }\href {https://doi.org/10.1364/AO.59.001363}
  {\bibfield  {journal} {\bibinfo  {journal} {Appl. Opt.}\ }\textbf {\bibinfo
  {volume} {59}},\ \bibinfo {pages} {1363} (\bibinfo {year}
  {2020})}\BibitemShut {NoStop}%
\bibitem [{\citenamefont {Beaurepaire}\ \emph {et~al.}(2015)\citenamefont
  {Beaurepaire}, \citenamefont {Vernier}, \citenamefont {Bocoum}, \citenamefont
  {B\"ohle}, \citenamefont {Jullien}, \citenamefont {Rousseau}, \citenamefont
  {Lefrou}, \citenamefont {Douillet}, \citenamefont {Iaquaniello},
  \citenamefont {Lopez-Martens}, \citenamefont {Lifschitz},\ and\ \citenamefont
  {Faure}}]{Beaurepaire2015}%
  \BibitemOpen
  \bibfield  {author} {\bibinfo {author} {\bibfnamefont {B.}~\bibnamefont
  {Beaurepaire}}, \bibinfo {author} {\bibfnamefont {A.}~\bibnamefont
  {Vernier}}, \bibinfo {author} {\bibfnamefont {M.}~\bibnamefont {Bocoum}},
  \bibinfo {author} {\bibfnamefont {F.}~\bibnamefont {B\"ohle}}, \bibinfo
  {author} {\bibfnamefont {A.}~\bibnamefont {Jullien}}, \bibinfo {author}
  {\bibfnamefont {J.-P.}\ \bibnamefont {Rousseau}}, \bibinfo {author}
  {\bibfnamefont {T.}~\bibnamefont {Lefrou}}, \bibinfo {author} {\bibfnamefont
  {D.}~\bibnamefont {Douillet}}, \bibinfo {author} {\bibfnamefont
  {G.}~\bibnamefont {Iaquaniello}}, \bibinfo {author} {\bibfnamefont
  {R.}~\bibnamefont {Lopez-Martens}}, \bibinfo {author} {\bibfnamefont
  {A.}~\bibnamefont {Lifschitz}},\ and\ \bibinfo {author} {\bibfnamefont
  {J.}~\bibnamefont {Faure}},\ }\href
  {https://doi.org/10.1103/PhysRevX.5.031012} {\bibfield  {journal} {\bibinfo
  {journal} {Phys. Rev. X}\ }\textbf {\bibinfo {volume} {5}},\ \bibinfo {pages}
  {031012} (\bibinfo {year} {2015})}\BibitemShut {NoStop}%
\bibitem [{\citenamefont {Ferri}\ \emph {et~al.}(2016)\citenamefont {Ferri},
  \citenamefont {Davoine}, \citenamefont {Fourmaux}, \citenamefont {Kieffer},
  \citenamefont {Corde}, \citenamefont {Phuoc},\ and\ \citenamefont
  {Lifschitz}}]{ferri2016effect}%
  \BibitemOpen
  \bibfield  {author} {\bibinfo {author} {\bibfnamefont {J.}~\bibnamefont
  {Ferri}}, \bibinfo {author} {\bibfnamefont {X.}~\bibnamefont {Davoine}},
  \bibinfo {author} {\bibfnamefont {S.}~\bibnamefont {Fourmaux}}, \bibinfo
  {author} {\bibfnamefont {J.}~\bibnamefont {Kieffer}}, \bibinfo {author}
  {\bibfnamefont {S.}~\bibnamefont {Corde}}, \bibinfo {author} {\bibfnamefont
  {K.~T.}\ \bibnamefont {Phuoc}},\ and\ \bibinfo {author} {\bibfnamefont
  {A.}~\bibnamefont {Lifschitz}},\ }\href@noop {} {\bibfield  {journal}
  {\bibinfo  {journal} {Scientific reports}\ }\textbf {\bibinfo {volume} {6}},\
  \bibinfo {pages} {1} (\bibinfo {year} {2016})}\BibitemShut {NoStop}%
\bibitem [{\citenamefont {Di~Piazza}\ \emph {et~al.}(2012)\citenamefont
  {Di~Piazza}, \citenamefont {M\"uller}, \citenamefont {Hatsagortsyan},\ and\
  \citenamefont {Keitel}}]{DiPiazza2012}%
  \BibitemOpen
  \bibfield  {author} {\bibinfo {author} {\bibfnamefont {A.}~\bibnamefont
  {Di~Piazza}}, \bibinfo {author} {\bibfnamefont {C.}~\bibnamefont {M\"uller}},
  \bibinfo {author} {\bibfnamefont {K.~Z.}\ \bibnamefont {Hatsagortsyan}},\
  and\ \bibinfo {author} {\bibfnamefont {C.~H.}\ \bibnamefont {Keitel}},\
  }\href {https://doi.org/10.1103/RevModPhys.84.1177} {\bibfield  {journal}
  {\bibinfo  {journal} {Rev. Mod. Phys.}\ }\textbf {\bibinfo {volume} {84}},\
  \bibinfo {pages} {1177} (\bibinfo {year} {2012})}\BibitemShut {NoStop}%
\bibitem [{\citenamefont {Blackburn}(2020)}]{Blackburn2020}%
  \BibitemOpen
  \bibfield  {author} {\bibinfo {author} {\bibfnamefont {T.~G.}\ \bibnamefont
  {Blackburn}},\ }\href {https://doi.org/10.1007/s41614-020-0042-0} {\bibfield
  {journal} {\bibinfo  {journal} {Reviews of Modern Plasma Physics}\ }\textbf
  {\bibinfo {volume} {4}},\ \bibinfo {pages} {5} (\bibinfo {year}
  {2020})}\BibitemShut {NoStop}%
\bibitem [{\citenamefont {Wang}\ \emph {et~al.}(2014)\citenamefont {Wang},
  \citenamefont {Liu}, \citenamefont {He}, \citenamefont {Pan}, \citenamefont
  {Zhou}, \citenamefont {Wu},\ and\ \citenamefont {Zhu}}]{Wang2014}%
  \BibitemOpen
  \bibfield  {author} {\bibinfo {author} {\bibfnamefont {H.}~\bibnamefont
  {Wang}}, \bibinfo {author} {\bibfnamefont {C.}~\bibnamefont {Liu}}, \bibinfo
  {author} {\bibfnamefont {X.}~\bibnamefont {He}}, \bibinfo {author}
  {\bibfnamefont {X.}~\bibnamefont {Pan}}, \bibinfo {author} {\bibfnamefont
  {S.}~\bibnamefont {Zhou}}, \bibinfo {author} {\bibfnamefont {R.}~\bibnamefont
  {Wu}},\ and\ \bibinfo {author} {\bibfnamefont {J.}~\bibnamefont {Zhu}},\
  }\href {https://doi.org/10.1017/hpl.2014.28} {\bibfield  {journal} {\bibinfo
  {journal} {High Power Laser Science and Engineering}\ }\textbf {\bibinfo
  {volume} {2}},\ \bibinfo {pages} {e25} (\bibinfo {year} {2014})}\BibitemShut
  {NoStop}%
\bibitem [{\citenamefont {Gerchberg}(1972)}]{gerchberg1972practical}%
  \BibitemOpen
  \bibfield  {author} {\bibinfo {author} {\bibfnamefont {R.~W.}\ \bibnamefont
  {Gerchberg}},\ }\href@noop {} {\bibfield  {journal} {\bibinfo  {journal}
  {Optik}\ }\textbf {\bibinfo {volume} {35}},\ \bibinfo {pages} {237} (\bibinfo
  {year} {1972})}\BibitemShut {NoStop}%
\bibitem [{\citenamefont {zhen Yang}\ \emph {et~al.}(1994)\citenamefont {zhen
  Yang}, \citenamefont {zhen Dong}, \citenamefont {yuan Gu}, \citenamefont {yao
  Zhuang},\ and\ \citenamefont {Ersoy}}]{Yang1994}%
  \BibitemOpen
  \bibfield  {author} {\bibinfo {author} {\bibfnamefont {G.}~\bibnamefont {zhen
  Yang}}, \bibinfo {author} {\bibfnamefont {B.}~\bibnamefont {zhen Dong}},
  \bibinfo {author} {\bibfnamefont {B.}~\bibnamefont {yuan Gu}}, \bibinfo
  {author} {\bibfnamefont {J.}~\bibnamefont {yao Zhuang}},\ and\ \bibinfo
  {author} {\bibfnamefont {O.~K.}\ \bibnamefont {Ersoy}},\ }\href
  {https://doi.org/10.1364/AO.33.000209} {\bibfield  {journal} {\bibinfo
  {journal} {Appl. Opt.}\ }\textbf {\bibinfo {volume} {33}},\ \bibinfo {pages}
  {209} (\bibinfo {year} {1994})}\BibitemShut {NoStop}%
\bibitem [{\citenamefont {Misell}(1973)}]{Misell1973}%
  \BibitemOpen
  \bibfield  {author} {\bibinfo {author} {\bibfnamefont {D.~L.}\ \bibnamefont
  {Misell}},\ }\href {https://doi.org/10.1088/0022-3727/6/1/102} {\bibfield
  {journal} {\bibinfo  {journal} {Journal of Physics D: Applied Physics}\
  }\textbf {\bibinfo {volume} {6}},\ \bibinfo {pages} {L6} (\bibinfo {year}
  {1973})}\BibitemShut {NoStop}%
\bibitem [{\citenamefont {Fienup}(1982)}]{Fienup1982}%
  \BibitemOpen
  \bibfield  {author} {\bibinfo {author} {\bibfnamefont {J.~R.}\ \bibnamefont
  {Fienup}},\ }\href {https://doi.org/10.1364/AO.21.002758} {\bibfield
  {journal} {\bibinfo  {journal} {Appl. Opt.}\ }\textbf {\bibinfo {volume}
  {21}},\ \bibinfo {pages} {2758} (\bibinfo {year} {1982})}\BibitemShut
  {NoStop}%
\bibitem [{\citenamefont {Zhou}\ \emph {et~al.}(2019)\citenamefont {Zhou},
  \citenamefont {Li}, \citenamefont {Liu},\ and\ \citenamefont
  {Su}}]{Zhou2019}%
  \BibitemOpen
  \bibfield  {author} {\bibinfo {author} {\bibfnamefont {P.}~\bibnamefont
  {Zhou}}, \bibinfo {author} {\bibfnamefont {Y.}~\bibnamefont {Li}}, \bibinfo
  {author} {\bibfnamefont {S.}~\bibnamefont {Liu}},\ and\ \bibinfo {author}
  {\bibfnamefont {Y.}~\bibnamefont {Su}},\ }\href
  {https://doi.org/10.1364/OE.27.008958} {\bibfield  {journal} {\bibinfo
  {journal} {Opt. Express}\ }\textbf {\bibinfo {volume} {27}},\ \bibinfo
  {pages} {8958} (\bibinfo {year} {2019})}\BibitemShut {NoStop}%
\bibitem [{\citenamefont {Antonello}\ and\ \citenamefont
  {Verhaegen}(2015)}]{Antonello2015}%
  \BibitemOpen
  \bibfield  {author} {\bibinfo {author} {\bibfnamefont {J.}~\bibnamefont
  {Antonello}}\ and\ \bibinfo {author} {\bibfnamefont {M.}~\bibnamefont
  {Verhaegen}},\ }\href {https://doi.org/10.1364/JOSAA.32.001160} {\bibfield
  {journal} {\bibinfo  {journal} {J. Opt. Soc. Am. A}\ }\textbf {\bibinfo
  {volume} {32}},\ \bibinfo {pages} {1160} (\bibinfo {year}
  {2015})}\BibitemShut {NoStop}%
\bibitem [{\citenamefont {Doelman}\ \emph {et~al.}(2018)\citenamefont
  {Doelman}, \citenamefont {Thao},\ and\ \citenamefont
  {Verhaegen}}]{Doelman2018}%
  \BibitemOpen
  \bibfield  {author} {\bibinfo {author} {\bibfnamefont {R.}~\bibnamefont
  {Doelman}}, \bibinfo {author} {\bibfnamefont {N.~H.}\ \bibnamefont {Thao}},\
  and\ \bibinfo {author} {\bibfnamefont {M.}~\bibnamefont {Verhaegen}},\ }\href
  {https://doi.org/10.1364/JOSAA.35.001410} {\bibfield  {journal} {\bibinfo
  {journal} {J. Opt. Soc. Am. A}\ }\textbf {\bibinfo {volume} {35}},\ \bibinfo
  {pages} {1410} (\bibinfo {year} {2018})}\BibitemShut {NoStop}%
\bibitem [{\citenamefont {Miao}\ \emph {et~al.}(2022)\citenamefont {Miao},
  \citenamefont {Feder}, \citenamefont {Shrock},\ and\ \citenamefont
  {Milchberg}}]{Miao22}%
  \BibitemOpen
  \bibfield  {author} {\bibinfo {author} {\bibfnamefont {B.}~\bibnamefont
  {Miao}}, \bibinfo {author} {\bibfnamefont {L.}~\bibnamefont {Feder}},
  \bibinfo {author} {\bibfnamefont {J.~E.}\ \bibnamefont {Shrock}},\ and\
  \bibinfo {author} {\bibfnamefont {H.~M.}\ \bibnamefont {Milchberg}},\ }\href
  {https://doi.org/10.1364/OE.454796} {\bibfield  {journal} {\bibinfo
  {journal} {Opt. Express}\ }\textbf {\bibinfo {volume} {30}},\ \bibinfo
  {pages} {11360} (\bibinfo {year} {2022})}\BibitemShut {NoStop}%
\bibitem [{\citenamefont {Weisse}\ \emph {et~al.}(2023)\citenamefont {Weisse},
  \citenamefont {Esslinger}, \citenamefont {Howard}, \citenamefont {Foerster},
  \citenamefont {Haberstroh}, \citenamefont {Doyle}, \citenamefont {Norreys},
  \citenamefont {Schreiber}, \citenamefont {Karsch},\ and\ \citenamefont
  {D\"{o}pp}}]{Weise2023}%
  \BibitemOpen
  \bibfield  {author} {\bibinfo {author} {\bibfnamefont {N.}~\bibnamefont
  {Weisse}}, \bibinfo {author} {\bibfnamefont {J.}~\bibnamefont {Esslinger}},
  \bibinfo {author} {\bibfnamefont {S.}~\bibnamefont {Howard}}, \bibinfo
  {author} {\bibfnamefont {F.~M.}\ \bibnamefont {Foerster}}, \bibinfo {author}
  {\bibfnamefont {F.}~\bibnamefont {Haberstroh}}, \bibinfo {author}
  {\bibfnamefont {L.}~\bibnamefont {Doyle}}, \bibinfo {author} {\bibfnamefont
  {P.}~\bibnamefont {Norreys}}, \bibinfo {author} {\bibfnamefont
  {J.}~\bibnamefont {Schreiber}}, \bibinfo {author} {\bibfnamefont
  {S.}~\bibnamefont {Karsch}},\ and\ \bibinfo {author} {\bibfnamefont
  {A.}~\bibnamefont {D\"{o}pp}},\ }\href {https://doi.org/10.1364/OE.483801}
  {\bibfield  {journal} {\bibinfo  {journal} {Opt. Express}\ }\textbf {\bibinfo
  {volume} {31}},\ \bibinfo {pages} {19733} (\bibinfo {year}
  {2023})}\BibitemShut {NoStop}%
\bibitem [{\citenamefont {Santarsiero}\ \emph {et~al.}(1999)\citenamefont
  {Santarsiero}, \citenamefont {Gori}, \citenamefont {Borghi},\ and\
  \citenamefont {Guattari}}]{Santarsiero1999}%
  \BibitemOpen
  \bibfield  {author} {\bibinfo {author} {\bibfnamefont {M.}~\bibnamefont
  {Santarsiero}}, \bibinfo {author} {\bibfnamefont {F.}~\bibnamefont {Gori}},
  \bibinfo {author} {\bibfnamefont {R.}~\bibnamefont {Borghi}},\ and\ \bibinfo
  {author} {\bibfnamefont {G.}~\bibnamefont {Guattari}},\ }\href
  {https://doi.org/10.1364/AO.38.005272} {\bibfield  {journal} {\bibinfo
  {journal} {Appl. Opt.}\ }\textbf {\bibinfo {volume} {38}},\ \bibinfo {pages}
  {5272} (\bibinfo {year} {1999})}\BibitemShut {NoStop}%
\bibitem [{\citenamefont {Alieva}\ and\ \citenamefont
  {Bastiaans}(2002)}]{Alieva2002}%
  \BibitemOpen
  \bibfield  {author} {\bibinfo {author} {\bibfnamefont {T.}~\bibnamefont
  {Alieva}}\ and\ \bibinfo {author} {\bibfnamefont {M.~J.}\ \bibnamefont
  {Bastiaans}},\ }\href {https://doi.org/10.1364/JOSAA.19.000481} {\bibfield
  {journal} {\bibinfo  {journal} {J. Opt. Soc. Am. A}\ }\textbf {\bibinfo
  {volume} {19}},\ \bibinfo {pages} {481} (\bibinfo {year} {2002})}\BibitemShut
  {NoStop}%
\bibitem [{\citenamefont {Ivanov}\ \emph {et~al.}(1992)\citenamefont {Ivanov},
  \citenamefont {Sivokon},\ and\ \citenamefont {Vorontsov}}]{ivanov1992phase}%
  \BibitemOpen
  \bibfield  {author} {\bibinfo {author} {\bibfnamefont {V.~Y.}\ \bibnamefont
  {Ivanov}}, \bibinfo {author} {\bibfnamefont {V.}~\bibnamefont {Sivokon}},\
  and\ \bibinfo {author} {\bibfnamefont {M.}~\bibnamefont {Vorontsov}},\
  }\href@noop {} {\bibfield  {journal} {\bibinfo  {journal} {JOSA A}\ }\textbf
  {\bibinfo {volume} {9}},\ \bibinfo {pages} {1515} (\bibinfo {year}
  {1992})}\BibitemShut {NoStop}%
\bibitem [{\citenamefont {Chessa}\ \emph {et~al.}(1999)\citenamefont {Chessa},
  \citenamefont {Galimberti}, \citenamefont {Barbini}, \citenamefont {Danson},
  \citenamefont {Giulietti}, \citenamefont {Giulietti},\ and\ \citenamefont
  {Gizzi}}]{chessa1999phase}%
  \BibitemOpen
  \bibfield  {author} {\bibinfo {author} {\bibfnamefont {P.}~\bibnamefont
  {Chessa}}, \bibinfo {author} {\bibfnamefont {M.}~\bibnamefont {Galimberti}},
  \bibinfo {author} {\bibfnamefont {A.}~\bibnamefont {Barbini}}, \bibinfo
  {author} {\bibfnamefont {C.}~\bibnamefont {Danson}}, \bibinfo {author}
  {\bibfnamefont {A.}~\bibnamefont {Giulietti}}, \bibinfo {author}
  {\bibfnamefont {D.}~\bibnamefont {Giulietti}},\ and\ \bibinfo {author}
  {\bibfnamefont {L.}~\bibnamefont {Gizzi}},\ }\href@noop {} {\bibfield
  {journal} {\bibinfo  {journal} {Laser and Particle Beams}\ }\textbf {\bibinfo
  {volume} {17}},\ \bibinfo {pages} {681} (\bibinfo {year} {1999})}\BibitemShut
  {NoStop}%
\bibitem [{\citenamefont {Tajima}\ and\ \citenamefont
  {Dawson}(1979)}]{TajimaDawson1979}%
  \BibitemOpen
  \bibfield  {author} {\bibinfo {author} {\bibfnamefont {T.}~\bibnamefont
  {Tajima}}\ and\ \bibinfo {author} {\bibfnamefont {J.~M.}\ \bibnamefont
  {Dawson}},\ }\href {https://doi.org/10.1103/PhysRevLett.43.267} {\bibfield
  {journal} {\bibinfo  {journal} {Phys. Rev. Lett.}\ }\textbf {\bibinfo
  {volume} {43}},\ \bibinfo {pages} {267} (\bibinfo {year} {1979})}\BibitemShut
  {NoStop}%
\bibitem [{\citenamefont {Esarey}\ \emph {et~al.}(2009)\citenamefont {Esarey},
  \citenamefont {Schroeder},\ and\ \citenamefont {Leemans}}]{Esarey2009}%
  \BibitemOpen
  \bibfield  {author} {\bibinfo {author} {\bibfnamefont {E.}~\bibnamefont
  {Esarey}}, \bibinfo {author} {\bibfnamefont {C.}~\bibnamefont {Schroeder}},\
  and\ \bibinfo {author} {\bibfnamefont {W.}~\bibnamefont {Leemans}},\
  }\href@noop {} {\bibfield  {journal} {\bibinfo  {journal} {Reviews of modern
  physics}\ }\textbf {\bibinfo {volume} {81}},\ \bibinfo {pages} {1229}
  (\bibinfo {year} {2009})}\BibitemShut {NoStop}%
\bibitem [{\citenamefont {Birdsall}\ and\ \citenamefont
  {Langdon}(2004)}]{BirdsallLangdon2004}%
  \BibitemOpen
  \bibfield  {author} {\bibinfo {author} {\bibfnamefont {C.~K.}\ \bibnamefont
  {Birdsall}}\ and\ \bibinfo {author} {\bibfnamefont {A.~B.}\ \bibnamefont
  {Langdon}},\ }\href@noop {} {\emph {\bibinfo {title} {Plasma Physics via
  Computer Simulation}}}\ (\bibinfo  {publisher} {Taylor and Francis Group},\
  \bibinfo {year} {2004})\BibitemShut {NoStop}%
\bibitem [{\citenamefont {Siegman}(1986)}]{Siegman86}%
  \BibitemOpen
  \bibfield  {author} {\bibinfo {author} {\bibfnamefont {A.~E.}\ \bibnamefont
  {Siegman}},\ }\href@noop {} {\emph {\bibinfo {title} {Lasers}}}\ (\bibinfo
  {publisher} {University Science Books},\ \bibinfo {year} {1986})\BibitemShut
  {NoStop}%
\bibitem [{\citenamefont {Pang}\ \emph {et~al.}(2017)\citenamefont {Pang},
  \citenamefont {Wang}, \citenamefont {Zhang}, \citenamefont {Cao},
  \citenamefont {Shi},\ and\ \citenamefont {Deng}}]{pang2017non}%
  \BibitemOpen
  \bibfield  {author} {\bibinfo {author} {\bibfnamefont {H.}~\bibnamefont
  {Pang}}, \bibinfo {author} {\bibfnamefont {J.}~\bibnamefont {Wang}}, \bibinfo
  {author} {\bibfnamefont {M.}~\bibnamefont {Zhang}}, \bibinfo {author}
  {\bibfnamefont {A.}~\bibnamefont {Cao}}, \bibinfo {author} {\bibfnamefont
  {L.}~\bibnamefont {Shi}},\ and\ \bibinfo {author} {\bibfnamefont
  {Q.}~\bibnamefont {Deng}},\ }\href@noop {} {\bibfield  {journal} {\bibinfo
  {journal} {Optics Express}\ }\textbf {\bibinfo {volume} {25}},\ \bibinfo
  {pages} {14323} (\bibinfo {year} {2017})}\BibitemShut {NoStop}%
\bibitem [{\citenamefont {Wu}\ \emph {et~al.}(2021)\citenamefont {Wu},
  \citenamefont {Wang}, \citenamefont {Chen}, \citenamefont {Liu},
  \citenamefont {Jin},\ and\ \citenamefont {Chen}}]{wu2021adaptive}%
  \BibitemOpen
  \bibfield  {author} {\bibinfo {author} {\bibfnamefont {Y.}~\bibnamefont
  {Wu}}, \bibinfo {author} {\bibfnamefont {J.}~\bibnamefont {Wang}}, \bibinfo
  {author} {\bibfnamefont {C.}~\bibnamefont {Chen}}, \bibinfo {author}
  {\bibfnamefont {C.-J.}\ \bibnamefont {Liu}}, \bibinfo {author} {\bibfnamefont
  {F.-M.}\ \bibnamefont {Jin}},\ and\ \bibinfo {author} {\bibfnamefont
  {N.}~\bibnamefont {Chen}},\ }\href@noop {} {\bibfield  {journal} {\bibinfo
  {journal} {Optics express}\ }\textbf {\bibinfo {volume} {29}},\ \bibinfo
  {pages} {1412} (\bibinfo {year} {2021})}\BibitemShut {NoStop}%
\bibitem [{\citenamefont {Frazier}(2018)}]{Frazier2018}%
  \BibitemOpen
  \bibfield  {author} {\bibinfo {author} {\bibfnamefont {P.~I.}\ \bibnamefont
  {Frazier}},\ }\href@noop {} {\bibinfo {title} {A tutorial on bayesian
  optimization}} (\bibinfo {year} {2018}),\ \Eprint
  {https://arxiv.org/abs/arXiv:1807.02811} {arXiv:1807.02811} \BibitemShut
  {NoStop}%
\bibitem [{\citenamefont {Williams}\ and\ \citenamefont
  {Rasmussen}(2006)}]{williams2006gaussian}%
  \BibitemOpen
  \bibfield  {author} {\bibinfo {author} {\bibfnamefont {C.~K.}\ \bibnamefont
  {Williams}}\ and\ \bibinfo {author} {\bibfnamefont {C.~E.}\ \bibnamefont
  {Rasmussen}},\ }\href@noop {} {\emph {\bibinfo {title} {Gaussian processes
  for machine learning}}},\ Vol.~\bibinfo {volume} {2}\ (\bibinfo  {publisher}
  {MIT press Cambridge, MA},\ \bibinfo {year} {2006})\BibitemShut {NoStop}%
\bibitem [{\citenamefont {Head}\ \emph {et~al.}(2020)\citenamefont {Head},
  \citenamefont {Kumar}, \citenamefont {Nahrstaedt}, \citenamefont {Louppe},\
  and\ \citenamefont {Shcherbatyi}}]{head2020scikit}%
  \BibitemOpen
  \bibfield  {author} {\bibinfo {author} {\bibfnamefont {T.}~\bibnamefont
  {Head}}, \bibinfo {author} {\bibfnamefont {M.}~\bibnamefont {Kumar}},
  \bibinfo {author} {\bibfnamefont {H.}~\bibnamefont {Nahrstaedt}}, \bibinfo
  {author} {\bibfnamefont {G.}~\bibnamefont {Louppe}},\ and\ \bibinfo {author}
  {\bibfnamefont {I.}~\bibnamefont {Shcherbatyi}},\ }\href@noop {} {\bibfield
  {journal} {\bibinfo  {journal} {Zenodo}\ } (\bibinfo {year}
  {2020})}\BibitemShut {NoStop}%
\bibitem [{\citenamefont {Shahriari}\ \emph {et~al.}(2015)\citenamefont
  {Shahriari}, \citenamefont {Swersky}, \citenamefont {Wang}, \citenamefont
  {Adams},\ and\ \citenamefont {De~Freitas}}]{shahriari2015taking}%
  \BibitemOpen
  \bibfield  {author} {\bibinfo {author} {\bibfnamefont {B.}~\bibnamefont
  {Shahriari}}, \bibinfo {author} {\bibfnamefont {K.}~\bibnamefont {Swersky}},
  \bibinfo {author} {\bibfnamefont {Z.}~\bibnamefont {Wang}}, \bibinfo {author}
  {\bibfnamefont {R.~P.}\ \bibnamefont {Adams}},\ and\ \bibinfo {author}
  {\bibfnamefont {N.}~\bibnamefont {De~Freitas}},\ }\href@noop {} {\bibfield
  {journal} {\bibinfo  {journal} {Proceedings of the IEEE}\ }\textbf {\bibinfo
  {volume} {104}},\ \bibinfo {pages} {148} (\bibinfo {year}
  {2015})}\BibitemShut {NoStop}%
\bibitem [{\citenamefont {Zalevsky}\ \emph {et~al.}(1996)\citenamefont
  {Zalevsky}, \citenamefont {Mendlovic},\ and\ \citenamefont
  {Dorsch}}]{zalevsky1996gerchberg}%
  \BibitemOpen
  \bibfield  {author} {\bibinfo {author} {\bibfnamefont {Z.}~\bibnamefont
  {Zalevsky}}, \bibinfo {author} {\bibfnamefont {D.}~\bibnamefont
  {Mendlovic}},\ and\ \bibinfo {author} {\bibfnamefont {R.~G.}\ \bibnamefont
  {Dorsch}},\ }\href@noop {} {\bibfield  {journal} {\bibinfo  {journal} {Optics
  Letters}\ }\textbf {\bibinfo {volume} {21}},\ \bibinfo {pages} {842}
  (\bibinfo {year} {1996})}\BibitemShut {NoStop}%
\end{thebibliography}%

\end{document}